\documentclass[iop,twocolappendix]{emulateapj}

\newcommand{\CIV}{\ion{C}{4}}

\newcommand{\HI}{\ion{H}{1}}
\newcommand{\NV}{\ion{N}{5}}

\newcommand{\OVI}{\ion{O}{6}}
\newcommand{\OVII}{\ion{O}{7}}
\newcommand{\OVIII}{\ion{O}{8}}
\newcommand{\SiIV}{\ion{Si}{4}}

\newcommand{\Halpha}{H$\alpha$}

\newcommand{\Msol}{\ensuremath{M_{\odot}}}
\newcommand{\Ne}{\ensuremath{n_{\mathrm{e}}}}

\newcommand{\nH}{\ensuremath{n_{\mathrm{H}}}}
\newcommand{\nHe}{\ensuremath{n_{\mathrm{He}}}}

\newcommand{\vLSR}{\ensuremath{v_\mathrm{LSR}}}

\newcommand{\nm}{\ensuremath{\mbox{\nm}}}
\newcommand{\cm}{\ensuremath{\mbox{cm}}}

\newcommand{\km}{\ensuremath{\mbox{km}}}

\newcommand{\pc}{\ensuremath{\mbox{pc}}}
\newcommand{\kpc}{\ensuremath{\mbox{kpc}}}

\newcommand{\s}{\ensuremath{\mbox{s}}}

\newcommand{\yr}{\ensuremath{\mbox{yr}}}
\newcommand{\Myr}{\ensuremath{\mbox{Myr}}}

\newcommand{\K}{\ensuremath{\mbox{K}}}

\newcommand{\pcc}{\ensuremath{\cm^{-3}}}
\newcommand{\pcmsq}{\ensuremath{\cm^{-2}}}

\newcommand{\ps}{\ensuremath{\s^{-1}}}

\newcommand{\pyr}{\ensuremath{\yr^{-1}}}

\newcommand{\kmps}{\km\ \ps}
\newcommand{\Msolpy}{\Msol\ \pyr}

\newcommand{\presalt}{\pcc\ \K}

\newcommand{\fuse}{\textit{FUSE}}

\newcommand{\rosat}{\textit{ROSAT}}

\newcommand{\citepsq}[1]{[\citealp{#1}]}
\newcommand{\citetsq}[1]{\citeauthor{#1} [\citeyear{#1}]}
\newcommand{\eqref}[1]{equation~(\ref{#1})}

\newcommand{\Astar}{A\tablenotemark{$\ast$}}
\newcommand{\Bext}{B$_\mathrm{ext}$}
\newcommand{\betaHVC}{\ensuremath{\beta_\mathrm{HVC}}}
\newcommand{\MdotHHVC}{\ensuremath{\dot{\mathcal{M}}_\mathrm{HVC}^\mathrm{H}}}
\newcommand{\MHVCinit}{\ensuremath{M_\mathrm{HVC,0}}}

\newcommand{\MHVCHIinit}{\ensuremath{M_\mathrm{HVC,0}^\mathrm{H\,I}}}
\newcommand{\MHVCHIlost}{\ensuremath{M_\mathrm{HVC,lost}^\mathrm{H\,I}}}
\newcommand{\MinitT}{\ensuremath{M_{\mathrm{init,}T}}}
\newcommand{\Mion}{\ensuremath{M_\mathrm{ion}}}

\newcommand{\Mionhalo}{\ensuremath{M_\mathrm{ion,halo}}}
\newcommand{\Mionave}{\ensuremath{\langle M_\mathrm{ion} \rangle_\mathrm{HVC}}}
\newcommand{\mion}{\ensuremath{m_\mathrm{ion}}}
\newcommand{\Nbarion}{\ensuremath{\bar{N}(\mathrm{ion})}}
\newcommand{\NbarCIV}{\ensuremath{\bar{N}(\mbox{\CIV})}}
\newcommand{\NbarNV}{\ensuremath{\bar{N}(\mbox{\NV})}}
\newcommand{\NbarOVI}{\ensuremath{\bar{N}(\mbox{\OVI})}}
\newcommand{\NCIV}{\ensuremath{N(\mbox{\CIV})}}
\newcommand{\NNV}{\ensuremath{N(\mbox{\NV})}}
\newcommand{\NOVI}{\ensuremath{N(\mbox{\OVI})}}
\newcommand{\NHVC}{\ensuremath{\mathcal{N}_\mathrm{HVC}}}
\newcommand{\nHcl}{\ensuremath{n_\mathrm{H,cl}}}
\newcommand{\RMW}{\ensuremath{R_\mathrm{MW}}}
\newcommand{\tf}{\ensuremath{t_\mathrm{f}}}
\newcommand{\THVC}{\ensuremath{T_\mathrm{HVC}}}
\newcommand{\Tsim}{\ensuremath{T_\mathrm{sim}}}
\newcommand{\vz}{\ensuremath{v_\mathrm{z}}}
\newcommand{\vzcl}{\ensuremath{v_\mathrm{z,cl}}}
\newcommand{\zmax}{\ensuremath{z_\mathrm{max}}}

\shorttitle{SIMULATIONS OF HVCS II}
\shorttitle{HENLEY ET AL.}

\begin{document}

\title{Simulations of High-Velocity Clouds. II. Ablation from High-Velocity Clouds as a Source of Low-Velocity High Ions}
\author{David B. Henley\altaffilmark{1}, Kyujin Kwak\altaffilmark{1,2}, and Robin L. Shelton\altaffilmark{1}}
\affil{$^1$Department of Physics and Astronomy, University of Georgia, Athens, GA 30602;\\
  dbh@physast.uga.edu, rls@physast.uga.edu \\
  $^2$Korea Astronomy and Space Science Institute (KASI), 776 Daedeok-daero, Yuseong-gu, Daejeon, 305-348, Republic of Korea;\\
  kkwak@kasi.re.kr}

\begin{abstract}
In order to determine if the material ablated from high-velocity clouds (HVCs) is a significant
source of low-velocity high ions (\CIV, \NV, and \OVI) such as those found in the Galactic halo, we
simulate the hydrodynamics of the gas and the time-dependent ionization evolution of its carbon,
nitrogen, and oxygen ions. Our suite of simulations examines the ablation of warm material from
clouds of various sizes, densities, and velocities as they pass through the hot Galactic halo. The
ablated material mixes with the environmental gas, producing an intermediate-temperature mixture
that is rich in high ions and that slows to the speed of the surrounding gas.  We find that the slow
mixed material is a significant source of the low-velocity \OVI\ that is observed in the halo, as it
can account for at least $\sim$1/3 of the observed \OVI\ column density. Hence, any complete model
of the high ions in the halo should include the contribution to the \OVI\ from ablated HVC
material. However, such material is unlikely to be a major source of the observed \CIV, presumably
because the observed \CIV\ is affected by photoionization, which our models do not include. We
discuss a composite model that includes contributions from HVCs, supernova remnants, a cooling
Galactic fountain, and photoionization by an external radiation field. By design, this model matches
the observed \OVI\ column density. This model can also account for most or all of the observed \CIV,
but only half of the observed \NV.
\end{abstract}

\keywords{
  Galaxy: halo ---
  hydrodynamics ---
  ISM: clouds ---
  methods: numerical ---
  ultraviolet: ISM}

\section{INTRODUCTION}
\label{sec:Introduction}

\setcounter{footnote}{2}

High ions from astrophysically abundant metals (e.g., \CIV, \NV, and \OVI) in the interstellar
medium (ISM) trace spatial and temporal transitions between the hot ($T \ga 10^6~\K$) and warm/cool
($T \la \mbox{few} \times 10^4~\K$) phases of the ISM. In the Galactic halo, above the disk, various
physical structures may give rise to such transitions, including radiatively cooling Galactic
fountain gas \citep{edgar86,shapiro93,benjamin93}, supernova remnants \citep{shelton98,shelton06},
evaporating cold clouds that are embedded in hot ambient gas \citep{bohringer87,borkowski90}, or
turbulent mixing layers formed where cool and hot gas move relative to one another
\citep{begelman90,slavin93a,esquivel06,kwak10}. Comparing the predictions of these various models
with observations of high ions provides information on which physical processes are important in the
Galaxy's halo.

There are plenty of observational data for such comparisons. High ions have been observed in the
Galactic halo via their far-ultraviolet absorption lines, both at low velocities ($\la$90~\kmps;
e.g.,
\citealt{pettini82,savage87,sembach92,savage97,savage00,savage03,zsargo03,indebetouw04b,savage09})
and high velocities ($\ga$90~\kmps; e.g.,
\citealt{murphy00,sembach00,sembach03,fox04,fox05,fox06,collins04,collins07,shull11}). The
low-velocity high ions' scale heights are $\approx$3--5~\kpc\ \citep{savage97,bowen08,savage09},
although the filling factor of the high-ion-rich material is small \citep[e.g.,][]{shelton07}.
\CIV\ and \OVI\ have also been observed in the halo via their emission lines
\citep{shelton01,shelton07,shelton10,korpela06,otte06,dixon06,welsh07,dixon08,park09}.

The observed ratios of high ion column densities are often used in attempts to distinguish between
models \citep[e.g.,][]{sembach92,spitzer96,sembach97,savage97,indebetouw04b}, leading to the
conclusion that a single type of model cannot explain all of the observations. However, another
important consideration is the quantities of each ion that the models predict. While the
normalization of many of the aforementioned models is essentially a free parameter, it is important
that the normalization required by the observations is physically reasonable (for example,
\citetsq{indebetouw04b} point out that a worrisomely large number of turbulent mixing layers is
needed to match the observed column densities). For other models, the normalization can be
constrained in advance, without reference to the column density measurements. For example, in the
model of \citet{shelton06}, an ensemble of supernova remnants (SNRs) above $|z| = 130~\pc$ can
account for 14--39\% of the typical latitude-corrected \OVI\ column density observed toward
extragalactic objects by \citet{savage03}.  In this case, the model's normalization is fixed using
independently determined values for the rate and scale height of Galactic supernovae.

Here, we consider a new model: high-velocity clouds (HVCs) interacting with their surroundings. HVCs
are interstellar clouds with $|\vLSR| \ga 90~\kmps$ \citep{wakker97}. The first HVCs were discovered
via their \HI\ 21-cm emission \citep{muller63}, but are now known also to have an ionized component
\citep[e.g.,][]{tufte04}.  On some sightlines, highly ionized high-velocity gas unassociated with
high-velocity \HI\ is observed \citep[e.g.,][]{sembach03}. HVCs may be material in a Galactic
fountain \citep[e.g.,][]{bregman80}, material stripped off satellite galaxies
\citep[e.g.,][]{gardiner96,putman04}, material falling into the Galaxy from extragalactic space
\citep[e.g.,][]{oort66,blitz99}, or material left over from the formation of the Galaxy
\citep{maller04}. High ions can arise from the turbulent mixing of cool cloud material with hot
ambient gas \citep{slavin93a,esquivel06,kwak10,kwak11}. There are constraints on the number of HVCs
in the Galactic halo. Given these constraints, we here consider whether or not material left behind
by HVCs is a significant source of the low-velocity high ions observed in the halo.

Our HVC simulations come from \citet[hereafter Paper~I]{kwak11}, which used hydrodynamical
simulations to study the evolution of initially spherical HVCs traveling through a hot ($10^6~\K$)
ambient medium, thought to be representative of the upper halo. Our assuming the presence of this
hot gas in the halo is consistent with observations of the diffuse soft X-ray background
\citep{burrows91,wang95,pietz98,wang98,snowden98,snowden00,kuntz00,smith07a,galeazzi07,henley08a,lei09,yoshino09,gupta09b,henley10b}
and of absorption lines, such as \OVII\ and \OVIII, in the X-ray spectra of active galactic nuclei
\citep{nicastro02,fang03,rasmussen03,mckernan04,williams05,fang06,bregman07,yao07a,yao08,yao09}.
The hot halo gas is likely due to Galactic fountains, with a possible contribution from accreting
extragalactic material \citep[and references therein]{henley10b}.  However, it should be noted that
the distribution of this gas is not currently well known, and it need not mostly fill the halo.

In Paper~I, we studied the high ions that arise when cool material ablates from an HVC and mixes
with the hot ambient gas. To this end, we self-consistently traced the non-equilibrium ionization
(NEI) evolution of carbon, nitrogen, and oxygen. In that paper, we concentrated on the high-velocity
ions, those with line-of-sight speeds $\ga90~\kmps$. We showed that the column densities and column
density ratios predicted by our models overlapped those observed toward Complex~C
\citep{sembach03,fox04,collins07}. However, the high-ion-bearing material in our simulations
eventually slows to ISM-like velocities, leading to copious quantities of low-velocity ions (with
$|v| \la 90~\kmps$; see Section~\ref{sec:Model} for the exact definition of ``low velocity'' used in
this paper). In this paper, we describe the evolution of these ions.  We use the Galactic HVC mass
infall rate to estimate the column densities of high ions in the halo expected from HVCs interacting
with hot ambient gas. When we compare these predictions with observations, we find that a
significant fraction ($\ga$30\%) of the observed low-velocity \OVI\ could be due to this process.
We therefore argue that any complete model of the high ions in the halo should take into account the
contribution due to HVCs.

The remainder of this paper is arranged as follows. In Section~\ref{sec:Model} we briefly describe
our hydrodynamical model (see Paper~I for more details). In Section~\ref{sec:Ions} we describe the
evolution of the low-velocity high ions, and how this is affected by our different model
parameters. In Section~\ref{sec:Observations} we present the average column densities of
low-velocity high ions predicted by our models, and compare them with observations. In
Section~\ref{sec:Profiles} we present \OVI\ column density profiles predicted by our models. We
discuss and summarize our results in Sections~\ref{sec:Discussion} and \ref{sec:Summary},
respectively.

\section{HYDRODYNAMICAL MODEL}
\label{sec:Model}

Our hydrodynamical model is described in full in Paper~I (see also \citealt{kwak10}). Here, we give
a brief overview of the model. The hydrodynamical simulations were carried out in 2D cylindrical
coordinates using FLASH version 2.5 \citep{fryxell00}. We tracked the ionization evolution of
carbon, nitrogen, and oxygen using the FLASH NEI module.\footnote{As noted in Paper~I, we do not
  include the important high ion \SiIV\ because it is more susceptible to photoionization. Modeling
  photoionization is beyond the scope of this paper.} The simulations include radiative cooling
using the default FLASH cooling curve, which is a piecewise power-law approximation of the
\citet{raymond77} cooling curve \citep{rosner78,peres82}. This cooling curve assumes collisional
ionization equilibrium (CIE); see \citet{kwak10} for some discussion of CIE versus NEI cooling
rates. Note that the model does not include a magnetic field nor thermal conduction (see
Section~\ref{subsubsec:NeglectedProcesses}).

In each simulation, the grid was initialized with a cool ($10^3$~K) HVC embedded in a hot ($10^6$~K)
ambient medium. All of the clouds are initially spherical. The density of hydrogen atoms in each
cloud (\nHcl\ in Table~\ref{tab:ModelParameters}) pertains to the density of the cloud
interior. Near the cloud's periphery, the density decreases smoothly to that of the ambient gas
($n_\mathrm{H,amb} = \nHcl/1000$), and the temperature increases smoothly from $10^3$ to $10^6$~K,
except in Model~E, which has a sharp edge (see Figure~1 in Paper~I). Initially, the cloud interior,
ambient medium, and transition zone are in pressure balance. Note that, although the densities are
expressed in terms of the hydrogen number density, the gas also includes helium ($\nHe / \nH =
0.1$), so the total number density of atoms and ions is $1.1\nH$. We assume that hydrogen and helium
are fully ionized at all temperatures, so the electron density $\Ne = 1.2 \nH$.

The simulations were carried out in the initial rest-frame of the HVC; i.e., the HVC was initially
at rest, while the ambient medium moved upward with velocity $|\vzcl|$, where \vzcl\ is the HVC's
initial velocity in the observer's frame (the observer is assumed to be located below the domain,
looking vertically upward). The boundary conditions allowed material to flow in from the bottom of
the domain and flow off the top of the domain. Before extracting masses or column densities of high
ions as a function of velocity, we transformed the velocities to the observer's frame.

\begin{deluxetable}{ccccc}
\tablewidth{0pt}
\tabletypesize{\footnotesize}
\tablecaption{Model Parameters\label{tab:ModelParameters}}
\tablehead{
\colhead{Model}         & \colhead{$r_0$}      & \colhead{\vzcl}   & \colhead{\nHcl}       & \colhead{\MHVCinit} \\
\colhead{}              & \colhead{(\pc)}       & \colhead{(\kmps)} & \colhead{(\pcc)}      & \colhead{(\Msol)} \\
\colhead{(1)}           & \colhead{(2)}         & \colhead{(3)}     & \colhead{(4)}         & \colhead{(5)}
}

\startdata
A &  20 & $-$100 & 0.1  & 120              \\
B & 150 & $-$100 & 0.1  & $4.9\times 10^4$ \\
C & 150 & $-$150 & 0.1  & $4.9\times 10^4$ \\
D & 150 & $-$300 & 0.1  & $4.9\times 10^4$ \\
E & 150 & $-$150 & 0.1  & $4.9\times 10^4$ \\
F & 300 & $-$100 & 0.1  & $4.0\times 10^5$ \\
G & 150 & $-$100 & 0.01 & $4.9\times 10^3$
\enddata

\tablecomments{
  Column~(1): Model identifiers.
  Column~(2): Initial cloud radius. For all except Model~E, this radius is approximate, as the clouds do not have sharp edges (see Figure~1 in Paper~I).
  Column~(3): Initial velocity of the cloud along the $z$-direction measured in the observer's frame.
  Column~(4): Initial hydrogen number density of the cloud at its center. The ambient number density is $1/1000$ this value.
  Column~(5): Initial total mass of cloud, including hydrogen and helium. For all except Model~E, this mass is approximate, as the clouds do not have sharp edges. Here, we use \MinitT\ from
  Paper~I -- the initial mass of the material with $T < 10^4~\K$.
}

\end{deluxetable}

The parameters for each model in our suite of models are listed in Table~\ref{tab:ModelParameters}.
Model~B is our reference model. The other models allow us to investigate the effects of cloud size
(Models~A, B, and F), cloud velocity (Models~B, C, and D), cloud density (Models~B and G), and cloud
density profile (Models~C and E).

In the following, we distinguish between high- and low-velocity ions using the same cuts in velocity
that we used in Paper~I. In Models~A, B, F, and G (in which the initial velocity of the cloud was
$\vzcl = -100~\kmps$), material with $\vz \le -80~\kmps$ is defined as high-velocity, while that
with $\vz > -80~\kmps$ is defined as low-velocity, where \vz\ is the $z$-velocity of the gas in the
observer's frame. For Models~C, D, and E (in which \vzcl\ was $-150$ or $-300~\kmps$), the velocity
cut is at $\vz = -100~\kmps$.

When calculating the quantities of high ions that result from the cloud-ISM interactions, we assumed
the \citet{wilms00} interstellar abundances for carbon, nitrogen, and oxygen. Note that in Paper~I
we used older abundances, from \citet{allen73}.

\section{EVOLUTION OF LOW-VELOCITY HIGH IONS}
\label{sec:Ions}

The hydrodynamical interaction between the model HVCs and the ambient gas is described in detail in
Paper~I. We begin this section by giving an overview of the processes. As a model HVC moves through
the ISM, material is ablated from the cloud. The cool ablated material mixes with the hot
ambient gas, creating gas of intermediate temperature ($T \sim (\mbox{1--3}) \times 10^5$~\K) in
which high ions are abundant. The temperature of this mixed gas is affected both by radiative
cooling and continued mixing with the hot ambient gas. The fractions of high ions in the mixed gas
increase both by ionization of the initially cool ablated material, and by recombination of the
initially hot ambient material. However, the fractions of all ions differ by varying degrees from
those expected from CIE, as changes in the ionization balance lag behind the changes in the gas
temperature that result from mixing and radiative cooling. Soon after material is torn from the
clouds, it becomes relatively rich in high ions. At that time, the material is traveling almost as
fast as the HVC, but as the ablated material continues to mix with the ambient medium, it
slows. This causes the velocities of the ions to tend toward that of the ISM, and the ablated
material to drift further from the cloud.

\begin{figure*}
\centering
\plottwo{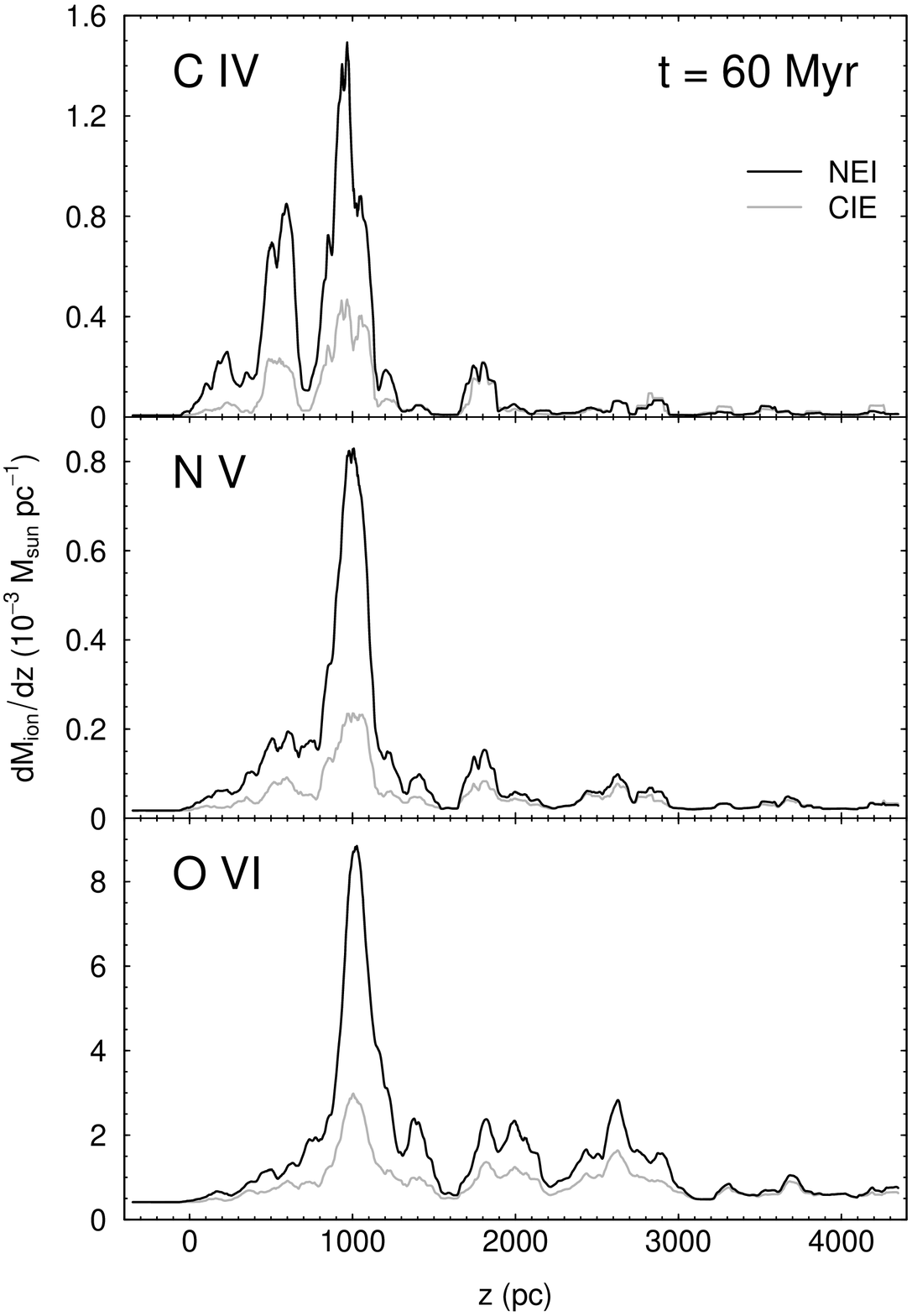}{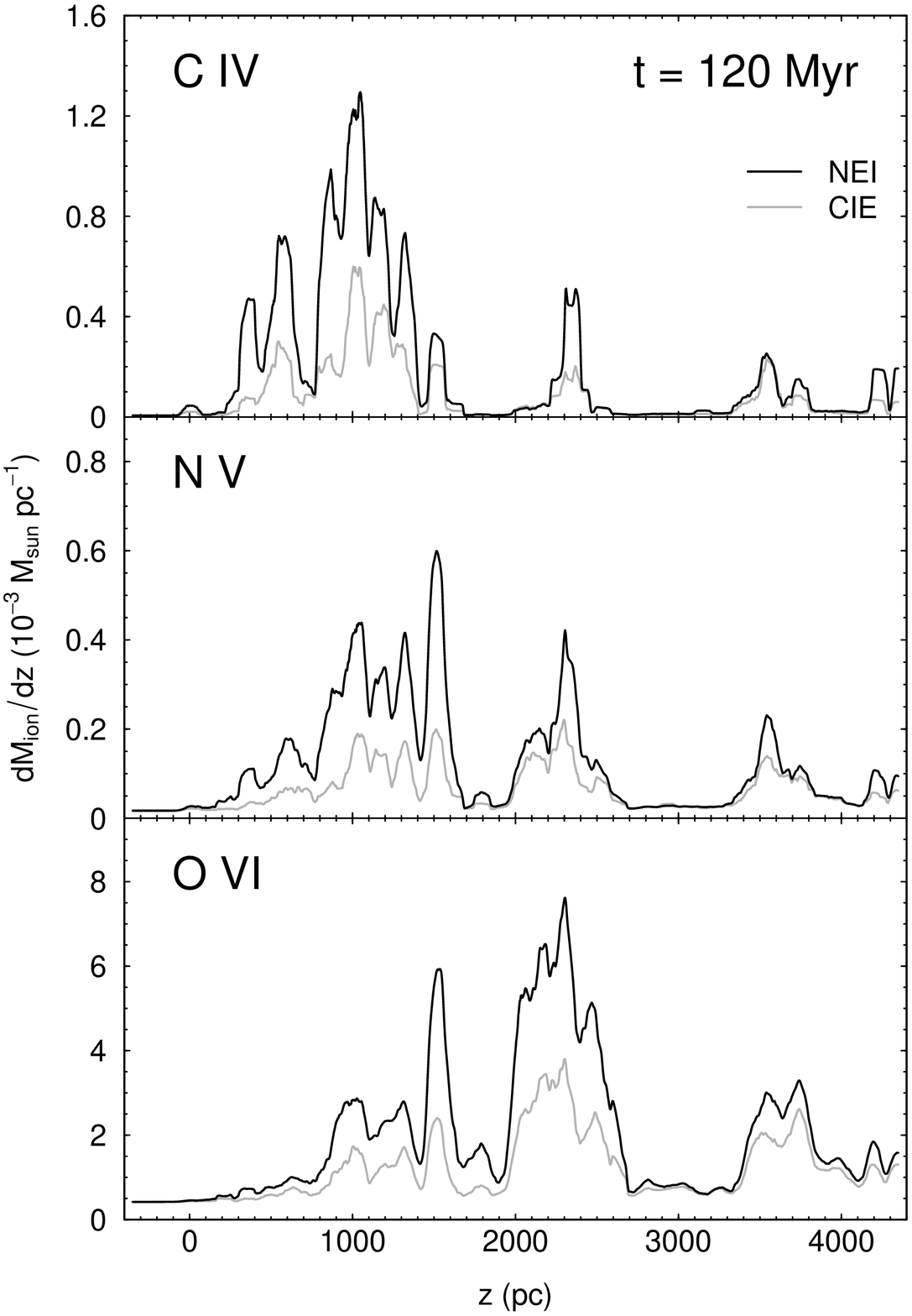}
\caption{Masses of low-velocity \CIV, \NV, and \OVI\ (top to bottom) as functions of height, $z$,
  measured in the frame in which the HVC was initially at rest at $z=0$. The data are from two
  epochs (60 and 120~\Myr) of an extended version of Model~B, Model~\Bext, in which the top of the
  domain is at $z=4400~\pc$ (instead of $z=2800~\pc$ as in the original Model~B). The mass of each ion
  was calculated by integrating over all cells at the specified height. The black and gray curves
  show NEI and CIE results, respectively. The data have been smoothed with a boxcar of full width
  100~\pc.
  \label{fig:IonMassVHeight}}
\end{figure*}

In Figure~\ref{fig:IonMassVHeight} we show where the low-velocity ions are located relative to the
HVC, by plotting the mass of each ion as a function of height. These plots were created using data
from a modified version of Model~B called Model~\Bext, in which the domain extends up to $\zmax =
4400~\pc$, instead of $\zmax = 2800~\pc$ as in the original Model~B (we increased \zmax\ for this
purpose so we could trace the high ions over a greater range of heights). The height, $z$, is
measured in the frame in which the HVC was initially at rest at $z=0$. The HVC shifts upward during
the course of the simulation, due to the ram pressure of the ambient medium. However, as this shift
is only $\sim$200~\pc\ by $t = 120~\Myr$, $z$ is approximately the height above the HVC's current
position.  We show results for two different epochs, $t = 60$ and 120~\Myr. In each panel, the black
curve shows the results of the NEI calculations, and gray curve shows the results obtained assuming
CIE, where the ion fractions depend only on the local gas temperature. Note that, in general, there
are significantly more high ions than are expected from CIE. Note also that the peaks seen at the
two epochs do not represent the same ions -- the peaks move through the domain at
$\sim$80-100~\kmps, and so would move $\sim$5--6~\kpc\ in 60~\Myr.

We see that the low velocity \CIV\ generally resides within 1400~\pc\ of the cloud (for both CIE and
NEI predictions) at both 60 and 120~\Myr. On average, both the \NV\ and the \OVI\ reside further
behind the cloud. This is as expected -- the mixed gas further behind the cloud contains a larger
fraction of initially ambient gas and so, ignoring the effects of radiative cooling for the moment,
is hotter and more highly ionized than that nearer to the cloud.

Eventually, near the top of the domain, the mixed gas generally starts becoming too hot even for
\OVI, due to the continued mixing of the ablated cloud material with the hot ambient gas. This
continued mixing and heating appears to be the ultimate fate for most of the ablated material -- by
the end of Model~B, 38\%\ of the material that was initially in the cloud and is now above $z =
2~\kpc$ is hotter than $6 \times 10^5~\K$, while only 13\%\ of this material is cooler than $1
\times 10^5~\K$. The fraction of the initial cloud material that is hotter than $6 \times 10^5~\K$
by the end of Model~B increases with increasing $z$. The hot mixed gas also becomes more tenuous,
because of the low density of the ambient gas. Both of these effects lead to a fall off in the
number of high ions. In addition, the mixed gas is closer to, although not yet in equilibrium.

Although most of the mixed gas is hot, it was noted in Paper~I that radiatively cooled mixed gas ($T
\la 10^4~\K$) accumulates along the symmetry ($r=0$) axis of the domain. This radiatively cooled gas
may be observable (e.g., via \Halpha\ emission), although such predictions are beyond the scope of
this paper. However, it should be noted that this accumulation of cooled gas along the symmetry axis
may be an artifact of the cylindrical geometry. Furthermore, because of the cylindrical geometry,
this cooled gas represents a small fraction of the mixed gas mass. This also means that any further
mixing of this cooled gas with hotter gas makes a negligible contribution to the total high ion
content.

It is useful to calculate the time evolution of the numbers of or masses of low-velocity high ions
that result from the cloud-ISM interaction. Summing the amounts of such ions in the model domain at
a given epoch provides one measure, but the masses of the ions in the domain are lower limits on the
true masses of the ions that result from the HVC-ISM interaction. This is because these masses do
not take into account material that has flowed out of the model domain (recall that each simulation
is carried out in the cloud's rest frame and that material is allowed to flow off the domain at the
top boundary).

For most models, we can make an estimate of the upper limit on a given ion's mass by including
material that has flowed off the domain. We are able to estimate the quantity of material that
leaves the domain between each epoch in the model (i.e., the times at which the hydrodynamical data
are output from the code) from the vertical velocity and position of the cells near the top of the
domain. Gas that is traveling upward at velocity $v$ and that is a distance $< v \Delta t$ from the
top of the domain at the current epoch will leave the grid by the next epoch, where $\Delta t$ is
the time between epochs (e.g., 0.5~\Myr\ for Model~B; note that $\Delta t$ does not refer to the
simulational timestep, which is much shorter). We record the mass of each high ion that passes
beyond the top of the domain at each epoch.

Clearly, we cannot trace the evolution of the escaping material after it has left the computational
domain, but we can tally the amount of a given high ion that has crossed the upper domain boundary at all
preceding times. This tally serves as an upper limit on the amount of that ion beyond the upper boundary
at any given time, because it ignores the possibility that some of the escaped material further ionizes
(as would occur when the gas mixes with additional hot ambient gas, raising the temperature above
that which is favorable to \CIV, \NV, and \OVI) or recombines. The sum of the amount of a given high
ion that has moved out of the domain and the amount currently in the domain forms our upper limit
on the total amount of that high ion present.

\begin{figure}
\plotone{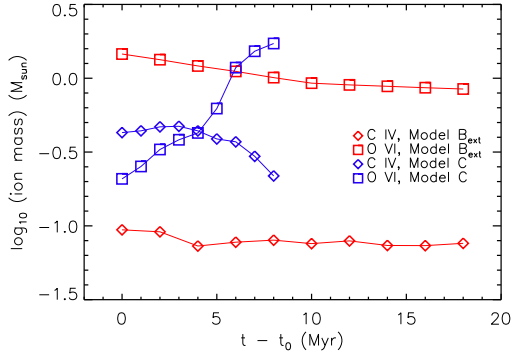}
\caption{Masses of low-velocity \CIV\ (diamonds) and \OVI\ (squares) in a single peak from the ion
  mass versus height distributions (Figure~\ref{fig:IonMassVHeight}) as functions of time. The
  Model~\Bext\ data (red) are from a peak that we traced from $t = t_0 = 72$ to $t = 90~\Myr$,
  during which time it moved from $z \approx 2000$ to $z \approx 3600~\pc$. The ion masses include
  all ions within $\Delta z = \pm 250~\pc$ of the peak.  The Model~C data (blue) are from a peak
  that we traced from $t = t_0 = 99$ to $t = 107~\Myr$, during which time it moved from $z \approx
  1300$ to $z \approx 2500~\pc$. The ion masses include all ions within $\Delta z = \pm 150~\pc$ of
  the peak.
  \label{fig:IonMassVTime}}
\end{figure}

Using Model~\Bext, we investigated the evolution of the low-velocity high ions after they have risen
above $z = 2800~\pc$, the maximum height in Model~B. We do this by following the time-evolution of
the ion masses contained in select peaks in the Model~\Bext\ mass-versus-height distribution (e.g.,
Figure~\ref{fig:IonMassVHeight}) as these peaks drift beyond the height of
Model~B. Figure~\ref{fig:IonMassVTime}, for example, shows the masses of \OVI\ and \CIV\ in a peak
that we traced from $t = 72$ to 90~\Myr, during which time it drifted from $z \approx 2000$ to
$\approx 3600~\pc$. Both the mass of \OVI\ and the mass of \CIV\ in the peak decreased during this
time (as did that of \NV, not shown).  Assuming that this trend applies to all of the low-velocity
high ions that rise past $z = 2800~\pc$ in Models~B and \Bext, the tally of all low-velocity high
ions that have passed the upper boundary in Model~B by some epoch of interest is an upper limit on
the mass of high ions above the boundary at that epoch. We expect this also to be true for the other
models with $\vzcl = -100~\kmps$ (Models~A, F, and G).

In models with $|\vzcl| > 100~\kmps$, the ram pressure of the ambient medium pushes the cloud upward
during the course of the simulation (see Figure~4 in Paper~I). As a result of this shift and the
greater flow speed in the ambient medium, in the later stages of the simulation ablated material
leaves the domain soon after it is torn from the HVC. The fractions of carbon, nitrogen, and oxygen
in the \CIV, \NV, and \OVI\ ionization levels, respectively, may still be increasing in this
outflowing material when it leaves the domain. Also, in Model~D in particular, a large fraction of
the high ions that flow off the domain do so at high velocities. This high-velocity material may
slow to low velocities after leaving the domain while remaining rich in high ions. Thus, it is
possible that the conclusion drawn from Model~\Bext, above, does not apply to Models~C, D, and E. We
found, for example, by examining a Model~C mass peak that the quantity of \OVI\ ions was still
increasing as the material approached the upper boundary of the domain (see
Figure~\ref{fig:IonMassVTime}), raising the possibility that it would have continued to increase
with time after crossing the $z = 2800~\pc$ mark, had the simulation domain included greater
heights. This was not the case for \CIV\ or \NV, whose masses had already begun to decrease while
the mass peak was still in the domain (see Figure~\ref{fig:IonMassVTime} for \CIV). Hence, for
Models~C, D, and E, taking into account material that has left the domain may not always yield upper
limits on the true total masses of the high ions.

\begin{figure}
\centering
\plotone{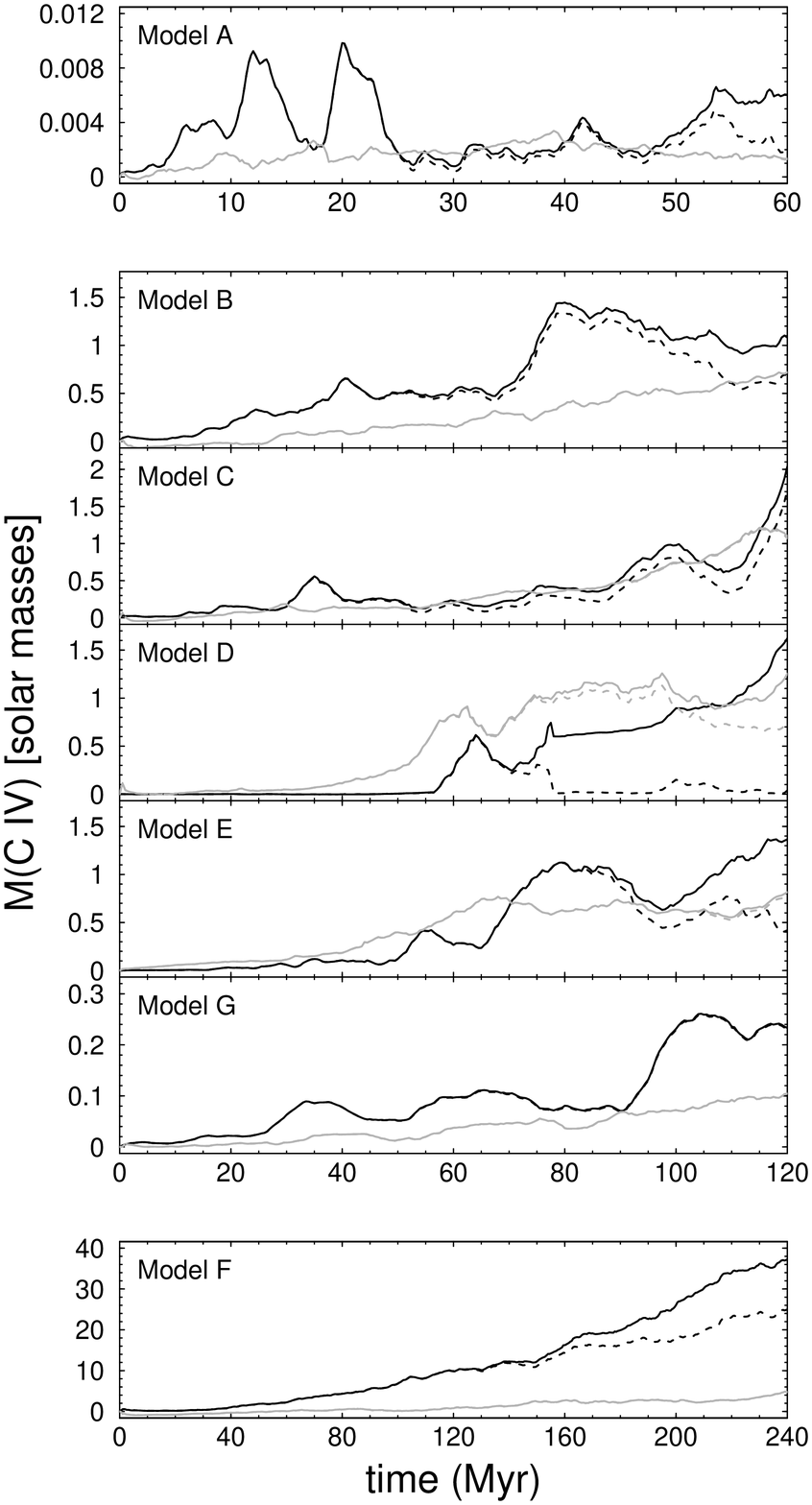}
\caption{Mass of \CIV\ as a function of time from our seven hydrodynamical models (note the
  different ranges on the time axes). In each panel, the black lines show the masses of low-velocity
  ions, and the gray lines show the masses of high-velocity ions. The dashed lines show the mass of
  each ion that is in the model domain, while the solid lines include ions that have escaped from
  the top of the domain (see text for details). Note that, in most panels, the gray solid and dashed
  lines are indistinguishable.
  \label{fig:C4Evolution}}
\end{figure}

\begin{figure}
\centering
\plotone{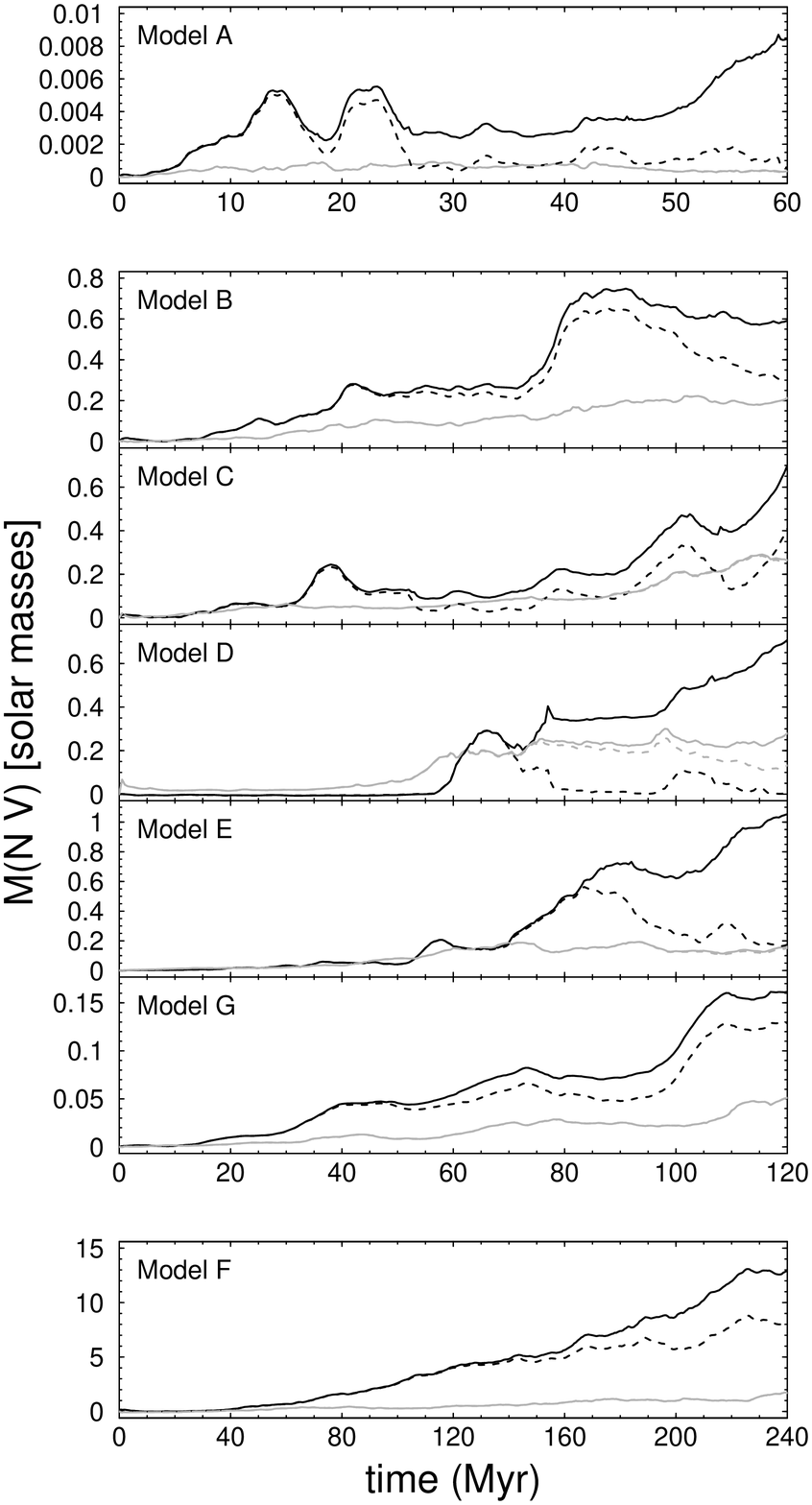}
\caption{As Figure~\ref{fig:C4Evolution}, but for \NV.
  \label{fig:N5Evolution}}
\end{figure}

\begin{figure}
\centering
\plotone{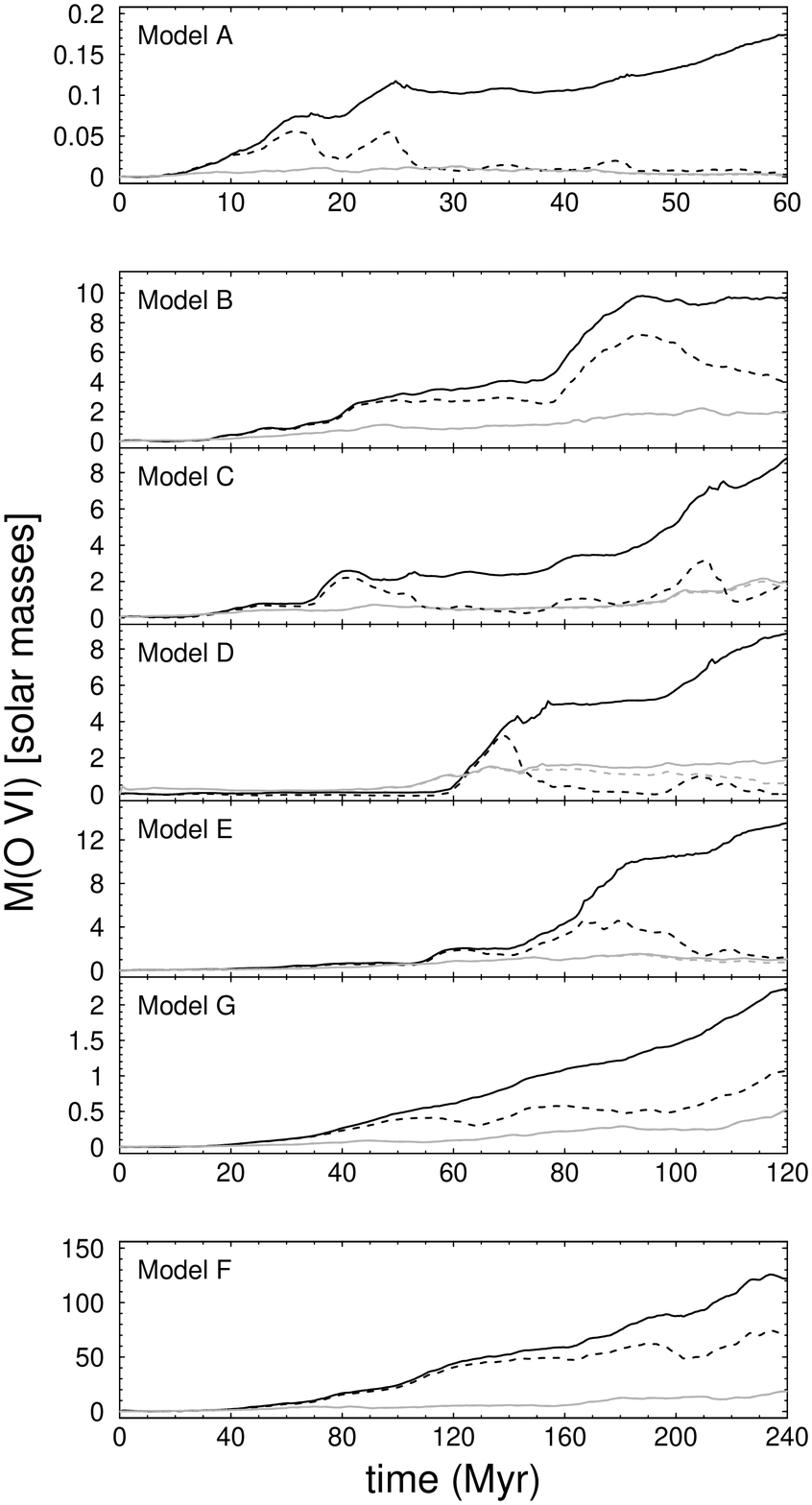}
\caption{As Figure~\ref{fig:C4Evolution}, but for \OVI.
  \label{fig:O6Evolution}}
\end{figure}

Figures~\ref{fig:C4Evolution} through \ref{fig:O6Evolution} show the lower and upper limits on the
masses of \CIV, \NV, and \OVI, respectively, that result from the HVC-ISM interaction. These masses
are plotted as functions of time for each of our seven hydrodynamical models.  We plot the masses of
both low-velocity and high-velocity ions (black and gray lines, respectively). In each case, we plot
the mass of the ion in the domain (dashed line), and that mass plus the mass of the ion in the all
material that has ever escaped from the domain (solid line). As noted previously, the former is the
lower limit on the true mass, while the latter is our best estimate of the upper limit.

\subsection{General Features in the Evolution of the Low-velocity High Ions}

Here, we discuss the evolution of the low-velocity high ions in our reference model, Model~B, and
point out features that are common across all or most of the other models. In the following
subsection, we will discuss how the different model parameters affect the evolution of these ions.

The number of low-velocity high ions generally increases with time throughout the simulations.
Admittedly, the masses of the high ions in the domain are decreasing toward the end of the Model~B
simulation. However, as $\sim$70\%\ of the cloud's initial \HI\ mass remains at the end of the
simulation (Paper~I), it is likely that this decrease is due to a relatively short-term fluctuation
in the number of high ions, rather than due to a long-term decline in the processes that lead to
high-ion-rich material. However, if we could run the simulation to much later times, we would
eventually exhaust the cloud, and the processes that lead to high-ion-rich material would decline
and eventually cease. The number of high ions would then decrease to zero as the high-ion-bearing
gas either radiatively cooled or mixed with the million-degree ambient gas.

When we compare the masses of low-velocity ions in Model~B that do and do not take into account
material that has escaped from the domain (solid and dashed black curves, respectively), we see that
these lines diverge by greater amounts as we go from \CIV\ to \NV\ to \OVI\
(Figures~\ref{fig:C4Evolution}, \ref{fig:N5Evolution}, and \ref{fig:O6Evolution},
respectively). This behavior is understandable from Figure~\ref{fig:IonMassVHeight}. The
low-velocity \CIV\ mainly resides close to the HVC, well away from the top of the domain, and so
very little low-velocity \CIV\ flows off the domain. In contrast, the low-velocity \OVI\ extends
further behind the HVC, nearer the top of the domain, and so significant quantities of this ion
escape from the domain. This general behavior is seen in the other models. However, it should be
noted that in Model~D the ambient material passes through the domain two or three times faster than
in the other models. As a result, the mixed \CIV-bearing gas is swept off the grid by the fast flow
of the ambient gas. The fact that significant quantities of low-velocity \CIV\ are lost from the
domain in this simulation is reflected by the wide gap between the solid and dashed black curves in
the Model~D panel of Figure~\ref{fig:C4Evolution}.

We note that, in most cases, most of the high ions are at low velocities, despite their origin in
the interaction of an HVC with its surroundings.  Although we are concentrating on the low-velocity
ions here, we point out that the solid gray (high-velocity material including that which has left
the domain) and dashed gray (high-velocity material in the domain) curves in
Figures~\ref{fig:C4Evolution} through \ref{fig:O6Evolution} are indistinguishable in nearly every
panel, implying that virtually no high-velocity high ions escape from the domain (i.e., the ions
slowed to low velocities before they left the domain).\footnote{We remind the reader that ``low-''
  and ``high-velocity'' refer to velocities in the observer's frame, whereas the simulations were
  carried out in the HVC's initial rest frame. Hence, ``low-velocity'' ions are moving quickly in
  the simulation domain, whereas ``high-velocity'' ions move relatively slowly.}  Again, the
exception to this is Model~D.

\subsection{Effects of the Cloud Velocity, Profile, Density, and Size}
\label{subsec:DifferentParameters}

\paragraph{Cloud Velocity}
Models~B, C, and D have identical initial clouds and ambient gas, but differing initial cloud
velocities ($-100$, $-150$, and $-300$~\kmps, respectively), allowing us to examine the effect of
cloud velocity. As noted in Paper~I, these velocities correspond to the subsonic, transonic, and
supersonic regimes, respectively, resulting in significant differences in the morphological
evolution of the HVCs.

One key difference between the models is that in Model~D a bow shock forms in front of the cloud,
which helps protect the cloud from ablation at earlier times, and delays the onset of mixing
(Paper~I). As a result, there are very few high ions (high- or low-velocity) in Model~D before $t
\sim 40~\Myr$. Another key difference is that the fast-moving ambient medium in Model~D tends to
sweep high ions off the domain before they are able to slow to low velocities in the observer's
frame. As result, although high-velocity ions start becoming abundant in Model~D at $t \sim
40~\Myr$, low-velocity ions do not become abundant until $t \sim 60~\Myr$. During this time delay,
the HVC and hence the material ablated from it decelerated sufficiently for the mixed, high-ion-rich
gas to reach low velocities before leaving the domain.

After $t \sim 60~\Myr$, despite the differences in the clouds' morphological evolution, the masses
of low-velocity high ions that include the ions that have escaped from the domain (solid black
curves in Figures~\ref{fig:C4Evolution} through \ref{fig:O6Evolution}) agree within a factor of
$\sim$3 for all three models.  If we consider only the ions in the domain (dashed black curves in
Figures~\ref{fig:C4Evolution} through \ref{fig:O6Evolution}), we see that Model~D has far fewer high
ions than Model~B after $t \sim 80~\Myr$. This is likely because the Model~D HVC shifts upward in
the domain during the course of the simulation, due to the large ram pressure of the ambient medium
(see Figure~4 in Paper~I). Eventually, the HVC gets so close to the top of the domain that, once
again, the high ions are unable to slow to low velocities before leaving the domain.

When considering the difference between the solid and dashed black curves in
Figures~\ref{fig:C4Evolution} through \ref{fig:O6Evolution}, Model~C generally lies between Models~B
and D. As in Model~D, the Model~C cloud shifts upward in the domain during the simulation. Because
the Model~C cloud gets closer to the top of the domain, relatively more low-velocity high ions are
swept off the top of domain than in Model~B. This has the greatest effect on \OVI, as \OVI\ tends to
exist further behind the HVC than \CIV\ or \NV. Model~E, in which the cloud velocity is the same as
in Model~C, is similarly affected.

\paragraph{Cloud Density Profile}
The only difference between the initial parameters in Models~C and E is the HVC's initial density
profile (smooth-edged versus sharp-edged; see Figure~1 in Paper~I). The large density contrast at
the edge of the Model~E cloud inhibits the growth of shear instabilities and delays the onset of
mixing, resulting in fewer high ions than in Model~C. However, after $t \sim 50~\Myr$, the
predictions from these models that include the ions that have escaped from the domain generally
agree within a factor of $\sim$2, with Model~E tending to yield larger masses.

\paragraph{Cloud Density}
The only differences between Models~B and G are that the cloud and ambient densities are 1/10 as
large in Model~G than in Model~B. As expected, the lower densities in Model G result in fewer high
ions. However, there are more high ions in Model~G than one would expect from a simple rescaling of
the Model~B results. This difference is related to the fact that radiative cooling operates at a
slower rate in Model~G, because of the lower density. This difference in cooling rate affects the
temperature distribution in the gas, which in turn affects the ionization and recombination
rates. Nevertheless, despite these differences, the ion masses in Models~B and G generally agree
within a factor of $\sim$2, after allowing for the factor of 10 difference in density.

\paragraph{Cloud Size}
Models~A, B, and F differ by the HVC's initial radius ($r_0 = 20$, 150, and 300~\pc, respectively)
and hence initial mass ($\MHVCinit = 120$, $4.9 \times 10^4$, and $4.0 \times 10^5$~\Msol,
respectively). Unsurprisingly, the more massive the HVC, the greater the mass of low-velocity ions
that results from its interaction with its surroundings.

We can examine this behavior more quantitatively by using a simple model of a uniform spherical HVC
shedding mass at a rate proportional to its surface area (Paper~I, Section~3.3.3). In this simple
model, we expect the mass of a given low-velocity ion, $\Mion$, at a rescaled simulation time of
$t/r_0$, to be proportional to the cloud's initial mass, \MHVCinit\ (see Equation~(\ref{eq:Mion2})
in the Appendix). The rescaling of the simulation time by dividing by $r_0$ is necessary if we wish
to compare models of different-sized HVCs at equivalent stages in their evolution, in the sense of
their having had the same fraction of their initial \HI\ mass ablated (see Paper~I). Note that, in
deriving $\Mion(t/r_0) \propto \MHVCinit$, we have ignored the fact that the ions of interest will
subsequently recombine or ionize. While this is a gross approximation, this simple model does
provide some insight into the relative behaviors of Models~A, B, and F.

We find that the scaling of \Mion\ with \MHVCinit\ from our hydrodynamical models is close to the
scaling expected from this simple model. When we compare Models~B and A, we find $\Mion(t/r_0)
\propto \MHVCinit^\gamma$, with $\gamma \approx 0.8$--1.0 (the exponent varies with $t/r_0$ and with
ion). When we compare Models~B and F, we find that $\gamma$ tends to be somewhat larger: $\gamma
\approx 0.9$--1.6. (In these comparisons, we used the ion masses that include material that has
escaped from the domain, and we ignored early times in the simulations [$t/r_0 <
  0.2~\Myr\ \pc^{-1}$, or $t < 4$, 30, and 60~\Myr\ for Models~A, B, and F, respectively].)  Thus,
the values of $\Mion(t/r_0)$ scale approximately as expected with \MHVCinit. The deviations from
$\Mion(t/r_0) \propto \MHVCinit$ may be due to our neglecting the recombination or ionization of the
high ions.

At later times in Model~A, the masses of low-velocity \NV\ and \OVI\ that do and do not take into
account material that has left the domain (black solid and black dashed lines, respectively) start
to diverge significantly. This divergence is due to an effect already mentioned for Models~C and E,
above: the cloud shifts upward in the model domain, leading to more low-velocity high ions being
swept off the top of domain. In the case of Model~A, it is the cloud's low mass--to--cross-section ratio
that makes it susceptible to being pushed upward by the ambient medium's ram pressure.

\section{COMPARISON WITH OBSERVATIONS: THE HALO COLUMN DENSITY OF LOW-VELOCITY HIGH IONS}
\label{sec:Observations}

\subsection{Model Column Density Predictions}
\label{subsec:ColDenPredictions}

In order to compare the results presented in the previous section with observations, we must
calculate from these results the expected column densities of low-velocity high ions. In order to do
this, we assume that the observed population of HVCs can be modeled by some number of model clouds,
each of which is similar to one of our simulated clouds. In this case,
\begin{equation}
  \NHVC = \frac{\MdotHHVC \THVC}{\MHVCHIinit}
\end{equation}
is the number of model HVCs needed to account for the mass of high-velocity material, where
\THVC\ and \MHVCHIinit\ are the lifetime and initial \HI\ mass\footnote{Note that the cloud masses
  in Table~\ref{tab:ModelParameters} are total masses, and so must be divided by 1.4 to give the
  \HI\ masses.} of the model cloud, respectively, and \MdotHHVC\ is the observed infall rate of HVCs
in the Galactic halo, expressed in terms of the total (neutral + ionized) hydrogen mass. By using an
observed HVC infall rate that includes both neutral and ionized material, we are implicitly assuming
that the observed HVCs began their lives as entirely neutral material, or, if they were initially
partially ionized, that this does not have a significant effect on the quantities of high ions that
result from their mixing with the hot halo.

If $\Mionave = \int \Mion(t) dt / \THVC$ is the time-averaged total mass of a given low-velocity ion
due to ablation from a single model HVC, then the total mass of that ion in the halo due to ablation
from the population of HVCs is given by
\begin{eqnarray}
  \Mionhalo &=& \Mionave \NHVC \nonumber \\
            &=& \frac{\MdotHHVC}{\MHVCHIinit} \int \Mion(t) dt.
  \label{eq:Mionhalo1}
\end{eqnarray}
If we assume that these ions are uniformly distributed in a cylindrical halo of radius \RMW\ above
and below the disk, the average column density for a vertical sightline is
\begin{equation}
  \Nbarion = \frac{\MdotHHVC}{2 \pi \RMW^2 \MHVCHIinit \mion} \int \Mion(t) dt,
  \label{eq:Nion}
\end{equation}
where \mion\ is the atomic mass of the ion (12.011, 14.00674, and 15.9994~u for \CIV, \NV, and \OVI,
respectively [taken from the FLASH code]).

The value of the HVC infall rate, \MdotHHVC, is uncertain. If the observed HVCs are accreting
extragalactic material, estimates for \MdotHHVC\ range from
$\sim$0.2~\Msolpy\ \citep{mirabel89,peek08} to $\sim$1~\Msolpy\ \citep{wakker97}. If the observed
HVCs are mainly due to a galactic fountain rather than due to accretion, there may be up to
$\sim$5~\Msolpy\ circulating through the halo \citep{wakker97}. We assume $\MdotHHVC =
0.5~\Msolpy$. Considering that infall rates for individual complexes are as large as
$\sim$0.1~\Msolpy\ for Complex~C \citep{wakker07,thom08}, $\ga$0.1~\Msolpy\ for the Smith Cloud
(using data from \citealt{lockman08} and \citealt{hill09}), and $\sim$0.1--0.5~\Msolpy\ for the
Magellanic Stream (the lower value is estimated from the mass of the negative-velocity portion of
the Stream from \citealt{bruns05}; the upper value is from \citealt{mirabel89}), our assumed value
of \MdotHHVC\ should be reasonably conservative. We also take $\RMW = 25~\kpc$ \citep{ferriere01}.

Ideally, the time integration of $\Mion(t)$ in Equation~(\ref{eq:Nion}) should be carried out over
the entire lifetime of the high ions in the halo. Note that this lifetime is, in general, not the
time it would take for the HVC to be disrupted in the halo, because (a) the low-velocity high ions
would tend to persist after the \HI\ HVC has been completely disrupted, or (b) the HVC may reach the
disk before it has been completely disrupted. In practice, if we integrate $\Mion(t)$ up to the end
of our simulations, then we obtain a lower limit on $\int \Mion(t) dt$, as we are neglecting the
contributions from beyond the ends of the simulations. In this subsection, we present the column
density predictions that result from the simulated part of the HVC evolution. In
Section~\ref{subsec:IonsAtLaterTimes}, below, we describe a method for including the estimated
contributions from beyond the ends of the simulations.

\begin{figure}
\centering
\plotone{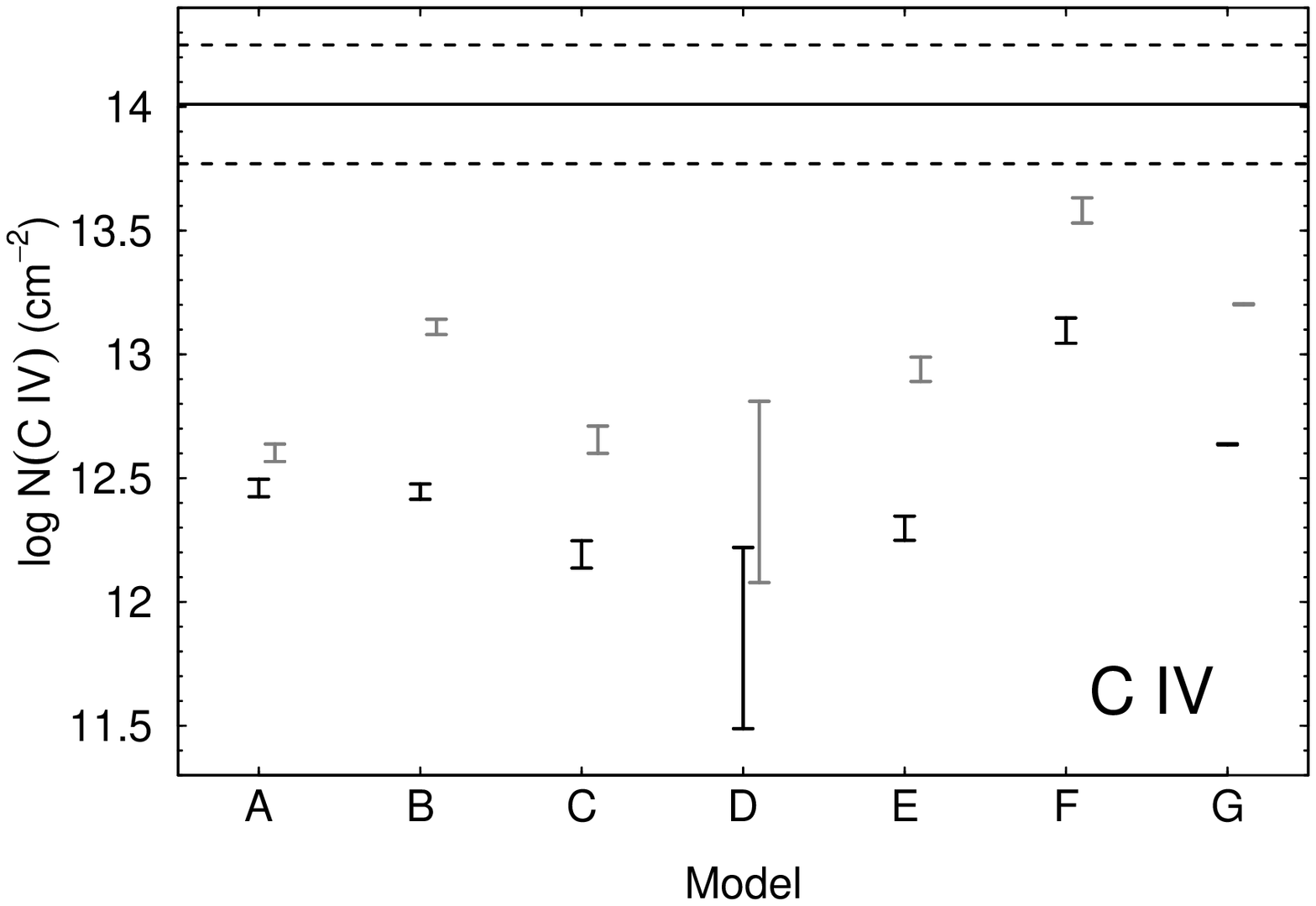}
\plotone{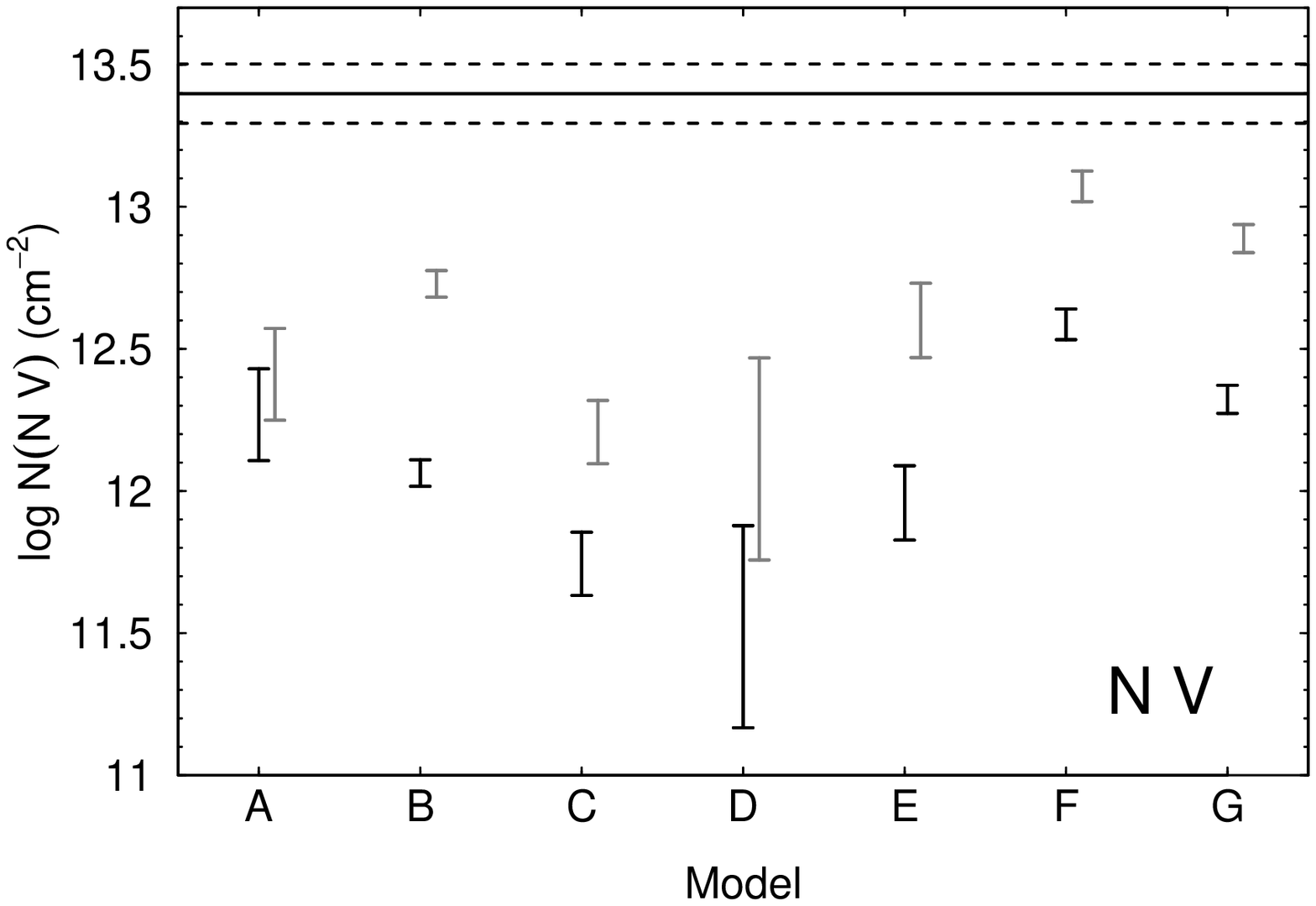}
\plotone{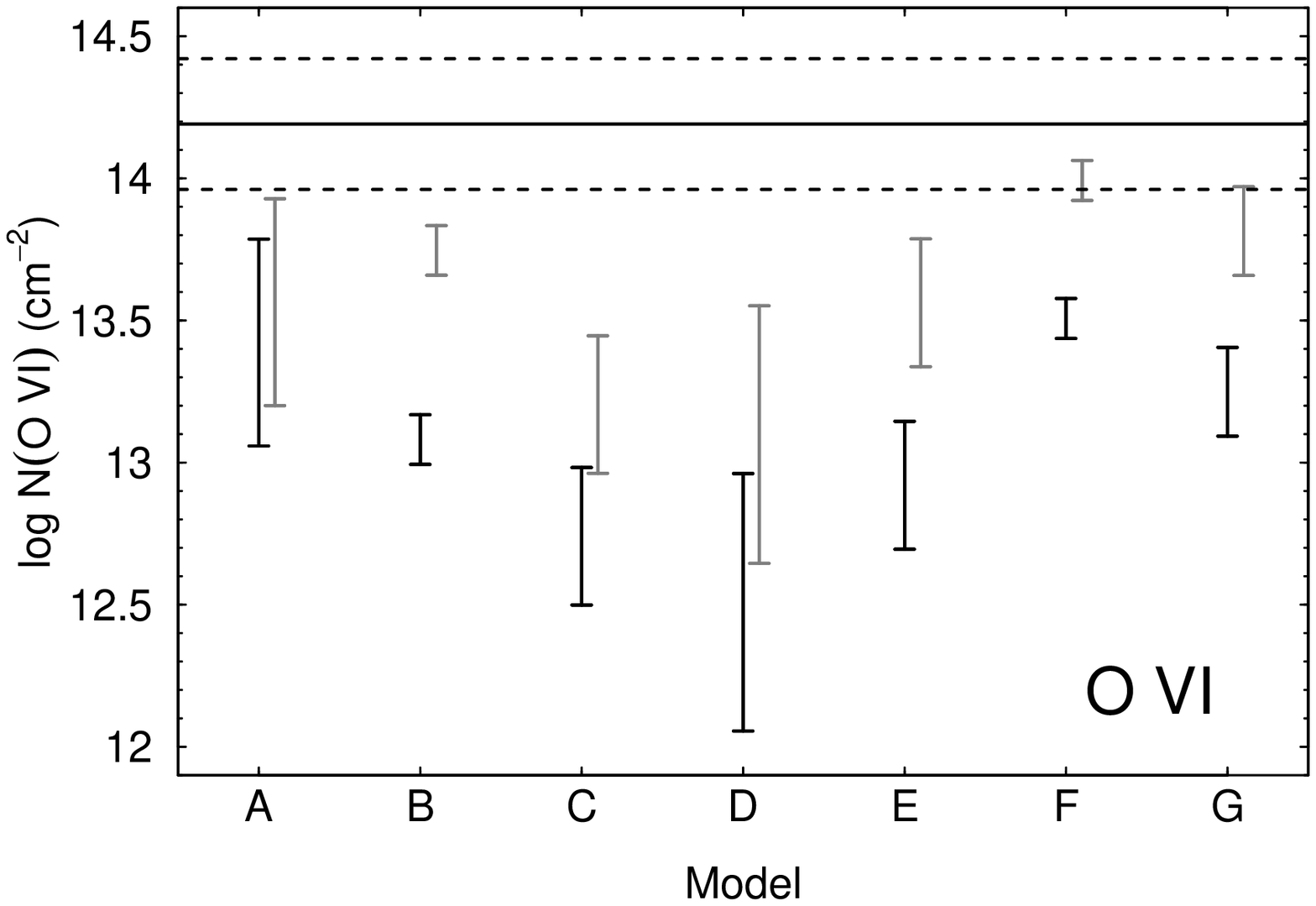}
\caption{Comparison of the observed and predicted low-velocity halo column densities for \CIV, \NV,
  and \OVI\ (top to bottom). The lower and upper ends of the black vertical bars show the ``Domain
  only'' and ``Domain + Escaped'' predictions, respectively, for each model from the upper half of
  Table~\ref{tab:ColDen}. For Model~A, we plot the value obtained by integrating $\Mion(t)$ up to
  the end of the simulation (i.e., A from Table~\ref{tab:ColDen}, rather than \Astar). These
  predicted column densities are lower limits, as we did not follow the high ions for the entire
  lifetime of each cloud. The gray vertical bars, offset slightly to the right, show the adjusted
  column density predictions from the lower half of Table~\ref{tab:ColDen}. These predictions take
  into account the estimated contribution from ions beyond the ends of the simulations (see
  Section~\ref{subsec:IonsAtLaterTimes} for details). The horizontal solid lines show the observed
  halo column densities. For \CIV\ we plot the average value of $\log [ \NCIV \sin |b| ]$, with the
  dashed lines indicating the standard deviation \citep[Table~4, specifically the values for all
    extragalactic sightlines]{savage09}. For \NV\ we plot the best-fit value of the midplane ion
  density multiplied by the ion scale height, with the dashed lines indicating the uncertainty
  \citep[Table~7, specifically the values that properly takes into account the fact that some
    sightlines yielded only upper limits on the \NV\ column density]{savage97}. We assume that the
  contribution from the Local Bubble is negligible for these ions. For \OVI\ we plot the average
  value of $\log [ \NOVI \sin |b| ]$, with the dashed lines indicating the standard deviation
  \citep[Table~3, specifically the values for the full dataset]{savage03}. We have subtracted $7
  \times 10^{12}~\pcmsq$ to allow for the contribution from the Local Bubble \citep{oegerle05}.
  \label{fig:Model+Obs}}
\end{figure}

The column densities calculated using Equation~(\ref{eq:Nion}) up to the times when the simulations
ended are shown in the upper half of Table~\ref{tab:ColDen}, and are plotted in black in
Figure~\ref{fig:Model+Obs}. In each case, we have calculated the column density using only the ions
that are in the simulational domain (``Domain only''), and using the ions in the domain plus those
that have escaped (``Domain + Escaped''). For Model~A, we also tabulate column densities resulting
from only the first 16~\Myr\ of the simulation (\Astar\ in Table~\ref{tab:ColDen}; see
Section~\ref{subsec:ModelComparison}).  Note in particular that the Domain-only values are strict
lower limits on the true model predictions (i.e., the column densities that the models would predict
if we could trace all of the high ions indefinitely). This is because the mass of an ion within the
simulation domain is a lower limit on the true value of $\Mion(t)$, and so each Domain-only column
density is calculated from the lower limit of the integral of a quantity that is itself a lower
limit.

\begin{deluxetable*}{cccccccccc}
\tablewidth{0pt}
\tabletypesize{\footnotesize}
\tablecaption{Average Column Densities Predicted for the Galactic Halo\label{tab:ColDen}}
\tablehead{
                   &                  & \multicolumn{2}{c}{\NbarCIV}         && \multicolumn{2}{c}{\NbarNV}          && \multicolumn{2}{c}{\NbarOVI} \\
                   &                  & \multicolumn{2}{c}{($10^{13}~\pcmsq$)}&& \multicolumn{2}{c}{($10^{13}~\pcmsq$)}&& \multicolumn{2}{c}{($10^{13}~\pcmsq$)} \\
\cline{3-4} \cline{6-7} \cline{9-10}
                   &                  & \colhead{Domain} & \colhead{Domain +}&& \colhead{Domain} & \colhead{Domain +}&& \colhead{Domain} & \colhead{Domain +}\\
\colhead{Model}    & \colhead{$\betaHVC$\tablenotemark{a}}
                                      & \colhead{only}   & \colhead{Escaped} && \colhead{only}   & \colhead{Escaped} && \colhead{only}   & \colhead{Escaped}
}
\startdata
A                  & \nodata          & 0.27             & 0.31              && 0.13             & 0.27              && 1.1              & 6.1 \\
\phantom{\tablenotemark{$\ast$}}A\tablenotemark{$\ast$}
                   & \nodata          & 0.088            & 0.088             && 0.043            & 0.044             && 0.35             & 0.41\\
B                  & \nodata          & 0.26             & 0.30              && 0.10             & 0.13              && 0.99             & 1.5 \\
C                  & \nodata          & 0.14             & 0.18              && 0.043            & 0.072             && 0.32             & 0.96\\
D                  & \nodata          & 0.031            & 0.17              && 0.015            & 0.076             && 0.11             & 0.91\\
E                  & \nodata          & 0.18             & 0.22              && 0.067            & 0.12              && 0.50             & 1.4 \\
F                  & \nodata          & 1.1              & 1.4               && 0.34             & 0.44              && 2.7              & 3.8 \\
G                  & \nodata          & 0.43             & 0.44              && 0.19             & 0.24              && 1.2              & 2.5 \\
\cutinhead{Adjusted Column Density Predictions (see Section~\ref{subsec:IonsAtLaterTimes})}
A                  & 0.721            & 0.37             & 0.43              && 0.18             & 0.37              && 1.6              & 8.5 \\
\phantom{\tablenotemark{$\ast$}}A\tablenotemark{$\ast$}
                   & \phantom{\tablenotemark{$\ast$}}0.272\tablenotemark{$\ast$}
                                      & 0.32             & 0.32              && 0.16             & 0.16              && 1.3              & 1.5 \\
B                  & 0.216            & 1.2              & 1.4               && 0.48             & 0.60              && 4.6              & 6.8 \\
C                  & 0.344            & 0.40             & 0.51              && 0.12             & 0.21              && 0.92             & 2.8 \\
D                  & 0.257            & 0.12             & 0.65              && 0.057            & 0.29              && 0.44             & 3.6 \\
E                  & 0.228            & 0.78             & 0.97              && 0.29             & 0.54              && 2.2              & 6.1 \\
F                  & 0.327            & 3.4              & 4.3               && 1.0              & 1.3               && 8.4              & 12  \\
G                  & 0.272            & 1.6              & 1.6               && 0.69             & 0.87              && 4.6              & 9.3
\enddata
\tablecomments{
These column densities were calculated using Equation~(\ref{eq:Nion}).
The column densities in the upper half of the table were calculated by integrating $\Mion(t)$ up to
the end of each simulation, except where noted. As a result, these column densities are lower
limits, because they neglect the contributions from ions at later times.
The adjusted column densities in the lower half of the table are those from the upper half divided
by $\betaHVC$, to take into account the estimated contribution from ions beyond the ends of the
simulations (see Section~\ref{subsec:IonsAtLaterTimes} for details).
For each ion, the first column gives the predicted column density calculated only from the ions in the domain.
The second column includes the ions that have escaped from the domain (see Section~\ref{sec:Ions}).
}
\tablenotetext{a}{The fraction of the HVC's initial \HI\ mass that has ablated and/or ionized
  by the end of the simulation (except where noted), from Paper~I. This quantity is not used in the upper half
  of the table, so we do not include it there.}
\tablenotetext{$\ast$}{Calculated by integrating $\Mion(t)$ to $t = 16~\Myr$, which corresponds to the
  same stage of evolution as the ends of Models~B and F. Similarly, $\betaHVC$ is taken at $t = 16~\Myr$,
  rather than at the end of the simulation.}
\end{deluxetable*}

\subsection{Model Comparison}
\label{subsec:ModelComparison}

With few exceptions, the Domain-only and Domain + Escaped predictions for each model agree with each
other within a factor of $\approx$2. The exceptions are all three ions in Model~D, and \OVI\ in
Models~A, C, and E.  In these cases, the HVCs shift upwards, which allows large numbers of
low-velocity high ions to escape from the domain during the course of the simulations.

Comparing models having different velocities (Models~B, C, and D), we find that their Domain +
Escaped predictions agree within a factor of 2, as do the Domain-only and the Domain + Escaped
predictions for models having different density profiles (Models~C and E). This is understandable,
given the similarities in the ion masses between these models (see
Section~\ref{subsec:DifferentParameters}). The Domain-only and the Domain + Escaped predictions for
models having different densities (Models~B and G) also agree within a factor of 2 (note that,
although $\Mion(t)$ in Equation~(\ref{eq:Nion}) is lower for Model~G, so too is \MHVCHIinit).

Comparing the results from the different-sized model clouds (Models~A, B, and F) is a little more
complicated, as the simulated portions of the clouds' lifetimes are not equivalent. As noted in
Section~\ref{subsec:DifferentParameters}, the simulation times must be divided by $r_0$ if we wish
to compare different-sized HVCs at equivalent stages in their evolution. Hence, we must reset the
end time for the $\Mion(t)$ integration, \tf, such that $\tf/r_0$ is the same for all clouds. We
achieve this by using the full simulation times as \tf\ for Models~B and F (120 and 240~\Myr,
respectively), and setting \tf\ to 16~\Myr\ for Model~A. The column density predictions resulting
from the new integration of the Model~A results are labeled \Astar\ in Table~\ref{tab:ColDen}.

Having obtained column density predictions for equivalent portions of the clouds' lifetimes, we can
compare them to the expectations for the simple model of a spherical cloud mentioned in
Section~\ref{subsec:DifferentParameters}. In this simple model, if $\tf/r_0$ is the same for all
clouds, we expect $\Nbarion \propto r_0$ (see Equation~(\ref{eq:Nion_r0}) in the Appendix).  The
$r_0$ ratios for Models~\Astar:B:F are 0.13:1:2 (Table~\ref{tab:ModelParameters}). In comparison,
the ratios of the Domain + Escaped column densities are 0.29:1:4.7 for \CIV, 0.34:1:3.4 for \NV, and
0.28:1:2.6 for \OVI. The ratios of the Domain-only column densities are generally similar:
0.34:1:4.3, 0.41:1:3.3, and 0.35:1:2.8 for \CIV, \NV, and \OVI, respectively.  Hence, the column
densities increase with $r_0$, but not linearly. The fact that the column densities for
Models~\Astar\ and F are somewhat larger relative to the Model~B values than expected from $r_0$ is
connected to the observation that the relationship between $\Mion(t/r_0)$ and \MHVCHIinit\ is
generally somewhat shallower than linear proportionality between Models~\Astar\ and B ($\gamma
\approx 0.8$--1.0), and generally somewhat steeper than linear proportionality between Models~B and
F ($\gamma \approx 0.9$--1.6; see Section~\ref{subsec:DifferentParameters}).

For Model~A, we can extend the integral of $\Mion(t)$ to later times, and naturally
\Nbarion\ increases as a result. However, we do not have data from Models~B and F for later times
with which to compare the Model~A results.

\subsection{Including the Contribution from Ions Beyond the Ends of the Simulations}
\label{subsec:IonsAtLaterTimes}

Here, we describe a method for estimating the contributions to the column densities from ions after
the simulations have ended. From Equation~(\ref{eq:Nion}), we see that the predicted column
densities depend on the integral $\int \Mion(t) dt$. Ideally, the time integration should be carried
out over the entire lifetime of the high ions in the halo, but in
Section~\ref{subsec:ColDenPredictions}, we carried out the integration only up to the end of each
simulation at time $t = \Tsim$.  To estimate the contribution from ions beyond the ends of the
simulations, we assume that each and every unit mass of \HI\ ``lost'' from the HVC to ablation and
ionization makes an approximately equal contribution to $\int \Mion(t) dt$. In this case, the
integral $\int \Mion(t) dt$ (integrated over all time) in Equation~(\ref{eq:Nion}) can be replaced
by $\int_{0}^{\Tsim} \Mion(t) dt / (\MHVCHIlost / \MHVCHIinit) \equiv \int_{0}^{\Tsim} \Mion(t) dt /
\betaHVC$, where \MHVCHIlost\ is the mass of \HI\ lost from the HVC to ablation and ionization, and
\betaHVC\ is the ratio of the lost mass to the HVC's initial \HI\ mass (defined in
Paper~I). Substituting this new integral into Equation~(\ref{eq:Nion}) allows us to approximate the
column densities that would be predicted if we could follow the full evolution of the cloud.  In
practice, this substitution can be accomplished by dividing the column densities calculated in
Section~\ref{subsec:ColDenPredictions} by $\betaHVC$.

The lower half of Table~\ref{tab:ColDen} contains the adjusted column density predictions for
Models~A through G, obtained by dividing the values in the upper half of the table by
$\betaHVC$. These adjusted column density predictions are also plotted in
Figure~\ref{fig:Model+Obs}, in gray. The adjustment increases the column densities for Models~B
through G by factors of $\sim$3--5 ($\sim$0.5--0.7~dex), and those for Model~A by $\approx$40\%
(0.14~dex). The lower half of Table~\ref{tab:ColDen} also contains the adjusted column densities for
Model~\Astar, obtained by dividing the relevant values in the upper half of the table by the value
of \betaHVC\ at $t=16~\Myr$.

After dividing by \betaHVC, the Domain-only column density predictions for Model~A are
systematically higher than those for Model~\Astar. As the Model~A predictions were obtained by
integrating the ion masses to a later time than the Model~\Astar\ predictions, this discrepancy
suggests that we are still underestimating the contribution from ions at times later than the ends
of the simulations. Nevertheless, the Domain-only predictions for Models~A and \Astar\ agree within
25\%, which gives us some confidence that dividing by \betaHVC\ leads to reasonable estimates of the
contribution of ions from beyond the ends of the simulations to the column densities. It should be
noted that the Domain + Escaped predictions for Models~A and \Astar\ in the lower half of
Table~\ref{tab:ColDen} are less consistent than the Domain-only predictions, especially for \NV\ and
\OVI. However, the Domain + Escaped predictions for Model~A may be significantly overestimated, as
relatively large quantities of low-velocity \NV\ and \OVI\ were swept off the top of the domain in
this simulation, and we were unable to follow these ions' subsequent evolution. Hence, the
discrepancy in the adjusted Domain + Escaped predictions for Models~A and \Astar\ does not argue
against the reliability of our method for estimating the contribution from ions beyond the ends of
the simulations.

It should be noted that there is a time delay between a unit mass of cool material being stripped
from the cloud and that material becoming rich in high ions. As a result, the above-described
division by $\betaHVC$ will neglect the contribution to $\int \Mion(t) dt$ from material that has
been stripped from the cloud, but has not yet completed its ionization evolution by the end of the
simulation. However, as this neglected contribution to $\int \Mion(t) dt$ is likely to be small, we do
not attempt to further adjust our column density predictions to take it into account.

\subsection{Comparison with Observations}
\label{subsec:ComparisonWithObservations}

Owing to the large number of models examined and the multiple methods for estimating their contributions
to the column densities of low-velocity high ions, our column density predictions cover a wide range.
At the high end, they can account for $\ga$1/3 of the observed \OVI\ column density, while our lower
limits account for only a few percent of the observed \CIV.

Let us first consider the column densities from the upper half of Table~\ref{tab:ColDen} (the values
that exclude the contributions from the times beyond the ends of the simulations). These are plotted
in black in Figure~\ref{fig:Model+Obs}, where they are compared with the observed values ($10 \times
10^{13}$, $2.5 \times 10^{13}$, and $15.5 \times 10^{13}~\pcmsq$ for \CIV, \NV, and \OVI,
respectively \citepsq{savage09,savage97,savage03}; see figure caption for details). In general, we
find that these model column densities account for only a small fraction of the low-velocity \CIV,
\NV, and \OVI\ observed in the halo. The exceptions are Models~A and F, whose Domain + Escaped
\OVI\ predictions agrees with the observed value within a factor of $\sim$2.5 and $\sim$4,
respectively.

When we include the estimated contribution from the times beyond the ends of the simulations
(tabulated in lower half of Table~\ref{tab:ColDen}, plotted in gray in Figure~\ref{fig:Model+Obs}),
we see that several models' predictions come within a factor of $\sim$3 of the observed \OVI\ column
density. The Domain-only and Domain + Escaped predictions from Models~B and G are within a factor of
3.5 of the observed value, while the Domain-only and Domain + Escaped predictions from Model~F and
the Domain + Escaped prediction from Model~A are within a factor of 2 of the observed value. The
Domain-only predictions are of particular interest, as they are lower limits on the true model
predictions. This is because the original Domain-only predictions are strict lower limits on the
true predictions (see Section~\ref{subsec:ColDenPredictions}), and dividing by \betaHVC\ may
underestimate the contribution from times beyond the ends of the simulations (see
Section~\ref{subsec:IonsAtLaterTimes}). Hence, our HVC models can account for a significant fraction
of the low-velocity \OVI\ observed in the halo (e.g., our reference model, Model~B, can account for
30--44\%\ of the observed \OVI, where the lower limit is taken from the Domain-only prediction, and
the upper limit is taken from the Domain + Escaped prediction).

For \CIV\ and \NV, we find that fewer of our models predict column densities that are within a
factor of $\sim$3 of the observed values. The Domain + Escaped \CIV\ and \NV\ predictions from
Model~F that include the contributions from times beyond the end of the simulation agree with the
observed values within factors of 3 and 2, respectively, while the equivalent Model~G
\NV\ prediction agrees with the observed value within a factor of 3. The other models typically
account for $\la$15\%\ and $\la$25\%\ of the observed \CIV\ and \NV, respectively (e.g., our
reference model, Model~B, can account for only 12--14\%\ of the observed \CIV, where the lower and
upper limits are again taken from the Domain-only and Domain + Escaped predictions,
respectively). We will discuss these results further in Section~\ref{sec:Discussion}

\section{COLUMN DENSITY PROFILES}
\label{sec:Profiles}

\begin{figure*}
\plottwo{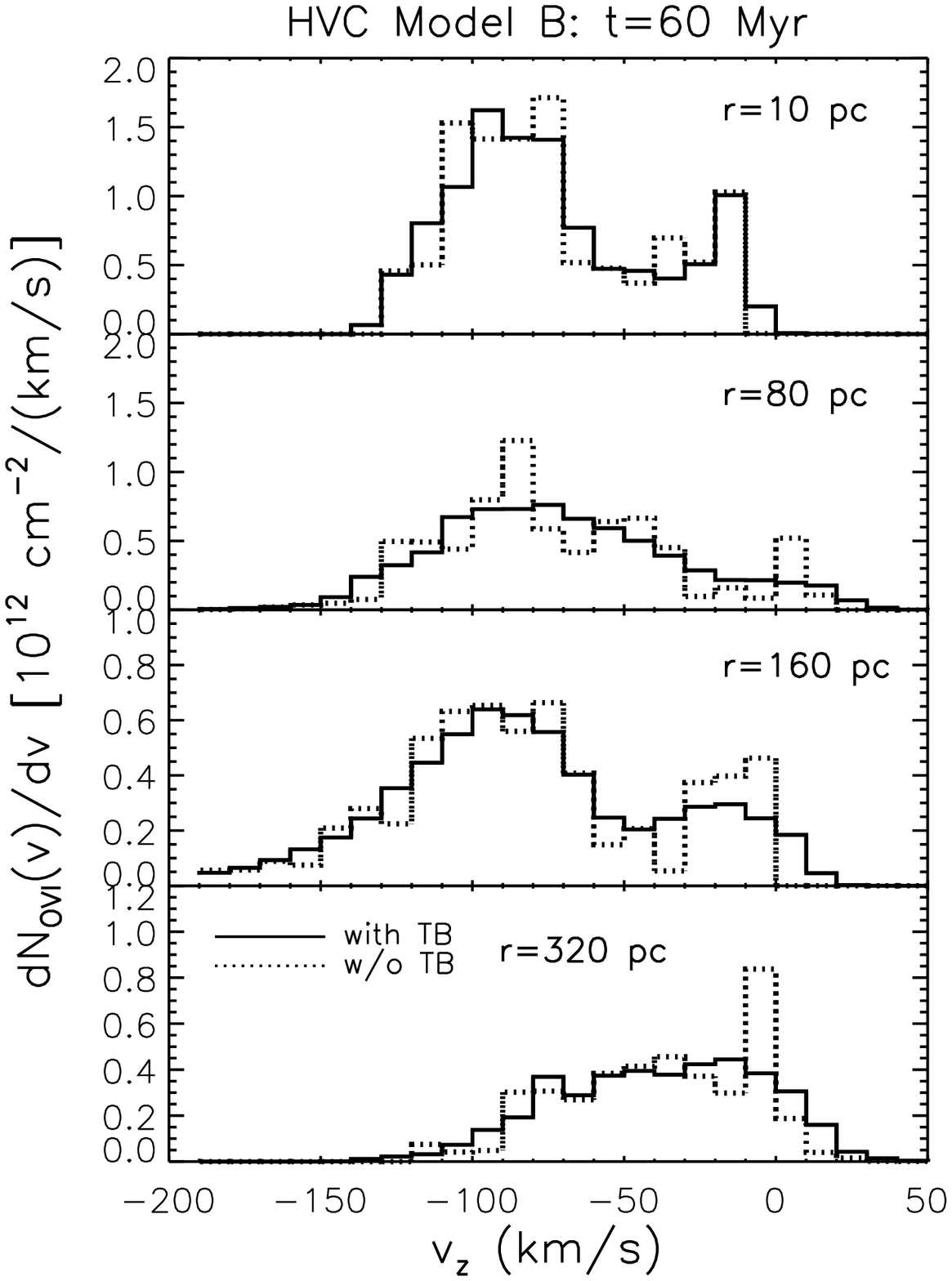}{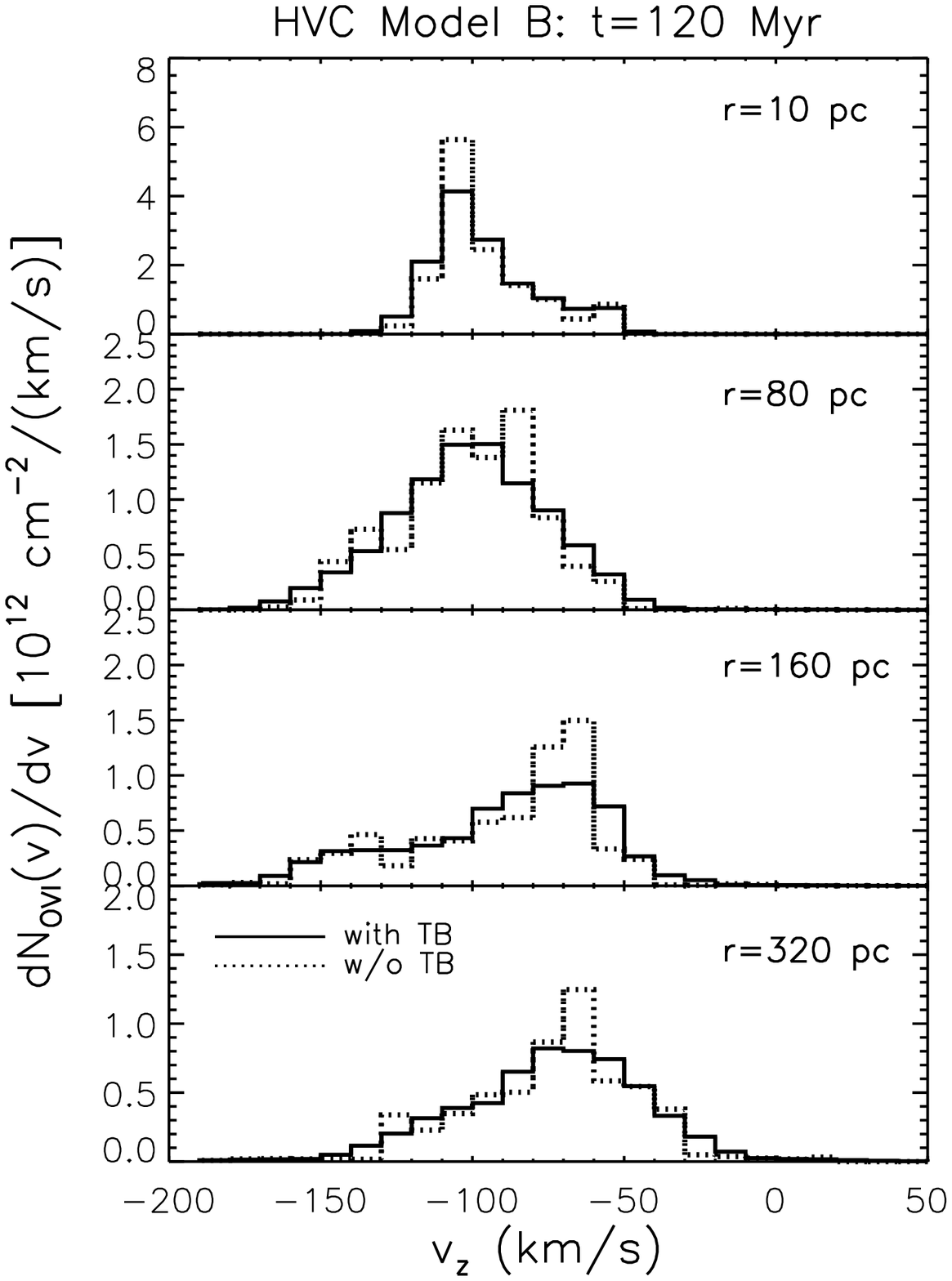}
\caption{\OVI\ column density profiles taken from two epochs (left: 60~\Myr, right: 120~\Myr) of Model~B.
  The column density profiles are for material lying along vertical sightlines at four different impact
  parameters relative to the cloud center (top to bottom: 10, 80, 160, and 320~\pc). The solid and dotted
  lines show the profiles with and without thermal broadening (see text for details).
  \label{fig:Profiles}}
\end{figure*}

\begin{figure}
\plotone{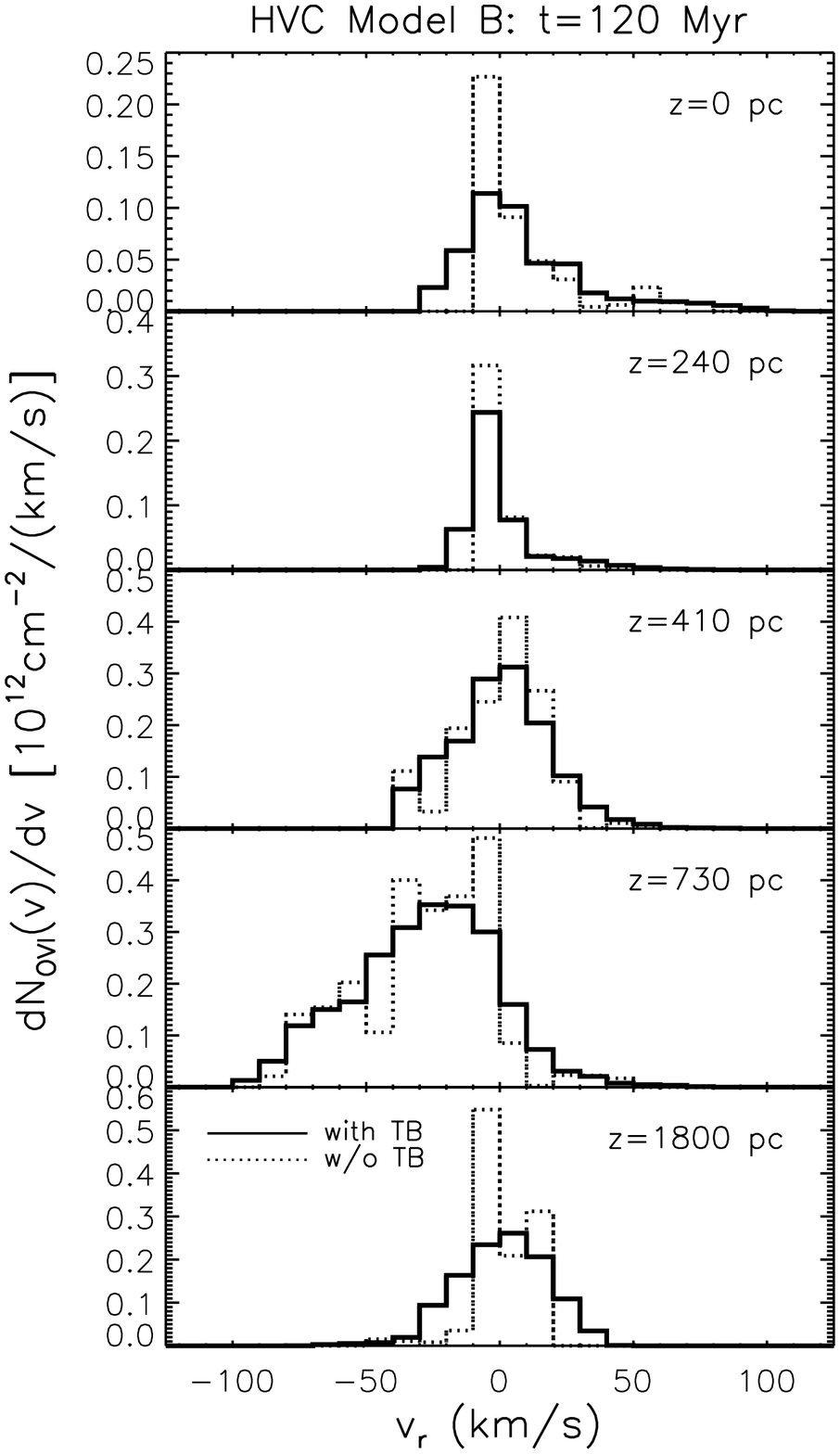}
\caption{As Figure~\ref{fig:Profiles}, but for horizontal sightlines through the Model~B domain at
  $t = 120~\Myr$. Profiles are plotted for sightlines at five different heights, measured relative
  to the cloud's initial position in the domain (top to bottom: 0, 240, 410, 730, and 1800~\pc).
  \label{fig:Profiles2}}
\end{figure}

In this section we present \OVI\ column density profiles (i.e., column density as a function of
velocity) derived from one of our HVC models, and compare them in general terms with the observed
profiles. Note that, although the main topic of this paper is low-velocity high ions, these column
density profiles span both low and high velocities.

Figure~\ref{fig:Profiles} shows \OVI\ profiles for four different vertical sightlines from two
epochs of Model~B, and Figure~\ref{fig:Profiles2} shows profiles for five different horizontal
sightlines from one epoch of the same model.  The column density of each hydrodynamical cell along
the line of sight was calculated from the gas density and the relevant ion fraction in that cell;
the velocity bin was chosen according to the cell's line-of-sight velocity, $v$, in the observer's frame
(for the vertical and horizontal sightlines, the observer is located below and to the right of the
model domain, respectively).  The dotted column density profiles in Figures~\ref{fig:Profiles} and
\ref{fig:Profiles2} do not include thermal broadening, and were constructed by summing the
contributions from all the cells along the line of sight, assuming that each cell's column density
profile is a Dirac $\delta$ function at $v$. The solid column density profiles do include thermal
broadening, which we approximated by convolving each cell's column density profile with a boxcar of
full width $2b$, where $b = (0.129~\kmps) \sqrt{(T/\K) / A}$ is the thermal velocity-spread
parameter, $T$ is the gas temperature in the cell, and $A$ is the relative atomic mass of the ion in
question \citep[Equation~3-21]{spitzer78}. Note that the profiles in Figures~\ref{fig:Profiles} and
\ref{fig:Profiles2} are much broader than the \fuse\ spectral resolution ($\sim$20~\kmps;
\citealt{sahnow00,wood02}).

Figures~\ref{fig:Profiles} and \ref{fig:Profiles2} show that the column density profiles with and
without thermal broadening are similar to each other, regardless of epoch or sightline. This means
that the broadening of the profiles is due to variation in the bulk fluid velocity along the line of
sight, rather than thermal broadening. For the horizontal sightlines, this variation in the bulk
fluid velocity is due to turbulence in the mixed gas. For the vertical sightlines, however, there is
an additional effect. The ions' velocities tend toward that of the ambient medium ($v=0$ in the
observer's frame) the further they are from the HVC. Therefore, a vertical sightline through our
model domains will typically sample both high- and low-velocity ions, spanning a continuous range of
velocities.  As a result of this additional effect, the column density profiles for the vertical
sightlines are generally somewhat broader than those for the horizontal sightlines.

\fuse\ observations of \OVI\ absorption in the halo show some qualitative agreement and some
qualitative disagreement with our predicted column density profiles. The observed line widths are
generally much broader than the expected thermal width, as in our model profiles. The observed
velocity-spread parameters for low- and high-velocity \OVI\ absorption are typically $b \sim 50$--70
and $\sim$30--50~\kmps, respectively \citep{savage03,sembach03}, compared with $b = 18~\kmps$ for
thermal broadening at $T = 3 \times 10^5~\K$. For several sightlines, absorption is observed over a
continuous range of velocities from $v \sim 0$ to $v \sim -200$ or +200~\kmps\ (see Figure~1 in
\citealt{wakker03}), in qualitative agreement with our model column density profiles for vertical
sightlines. However, the observed column density profiles typically have relatively more ions near
$v = 0$ than our model profiles. These extra ions near $v = 0$ are likely due to other sources of
low-velocity high ions, in addition to those predicted by our HVC models. We discuss further the
idea of multiple sources of high ions in the halo in Section~\ref{subsec:CompositeModel}.

\section{DISCUSSION}
\label{sec:Discussion}

We have analyzed existing simulations of HVCs (Paper~I) in order to estimate the quantities of
low-velocity ions that result from the passage of HVCs through the hot halo. In
Section~\ref{subsec:ComparisonWithObservations} we showed that some of our models could account for
a significant fraction of the observed low-velocity \OVI, but not of the low-velocity \CIV\ or \NV.
In this section, we first discuss our model assumptions, in particular neglected physical processes
(Section~\ref{subsubsec:NeglectedProcesses}), the pressure and temperature of the halo
(Section~\ref{subsubsec:PressureTemperatureHalo}), and the factors in Equation~(\ref{eq:Nion}) for
which we had to assume values (Section~\ref{subsubsec:ModelFactors}). Then, in
Section~\ref{subsec:CompositeModel}, we discuss our model alongside other models of the high ions in
the halo.

\subsection{Model Assumptions}

\subsubsection{Neglected Physical Processes}
\label{subsubsec:NeglectedProcesses}

Our model does not include a magnetic field, as our 2D geometry prevents our modeling realistic
field configurations. Magnetic fields are known to suppress the development of turbulence
\citep[e.g.,][]{ryu00}, while the development of turbulence differs in 2D and 3D
\citep[e.g.,][]{stone92d}. It is unclear whether adding a magnetic field and a third dimension to
the model would result in more or fewer high ions, compared to our current
simulations. \citet{kwak10} point out that their 2D hydrodynamic simulations of plane-parallel
mixing layers predict ion column density \textit{ratios} that are similar to those predicted by the
3D magnetohydrodynamic simulations of \citet{esquivel06}. However, we are unable to make a similar
statement regarding the magnitudes of the column densities.

Our model also does not include thermal conduction. Like a magnetic field, thermal conduction may
suppress the growth of instabilities \citep{orlando08}, although we cannot predict the extent to
which thermal conduction would affect the development of turbulence in our simulations. Assuming
that turbulence is able to develop, the inclusion of thermal conduction in our simulations would be
unlikely to greatly affect our results, as turbulent diffusion of heat should dominate over diffusion
by thermal conduction (see Section~5 of Paper~I).

\subsubsection{The Pressure and Temperature of the Halo}
\label{subsubsec:PressureTemperatureHalo}

Our model assumes that the halo pressure is uniform over the distance that the cloud falls through
the halo ($\ga$10~\kpc, assuming a cloud speed of 100~\kmps\ and a lifetime of $\ga$100~\Myr). This
would be an unrealistic assumption for HVCs traveling through the lower halo, where the total
interstellar pressure more than doubles from 3600 to 8700~\presalt\ between $|z| = 3$ and
1~\kpc\ \citep{ferriere98a}. However, the ambient halo pressure in our model is much lower than this
(230~\presalt\ in most models, 23~\presalt\ in Model~G), and may be more appropriate for HVCs
traveling through the upper halo ($|z| \ga 10~\kpc$). Although the pressure in the upper halo is
uncertain, in Paper~I we noted that our chosen ambient \textit{density} ($\nH = 1.0 \times
10^{-4}~\pcc$ for Models~A--F) is similar to some previous observational estimates of the density in
the upper halo ($\sim$$(\mbox{1--few}) \times 10^{-4}~\pcc$ for $|z| \ga 10~\kpc$; e.g.,
\citealt{weiner96,peek07,grcevich09}). As these observational estimates of the density in the upper
halo vary by only a factor of a few over a wide range of heights above the disk (tens of kpc), the
pressure gradients will not be large in the upper halo if the temperature is reasonably uniform in
this region.

The origin of HVCs remains uncertain, although \citet{peek07} point out that the observation of HVCs
$\ga$10~\kpc\ above the disk argues against their origin in a Galactic fountain
\citep{bregman80}. Some HVCs have low metallicities (e.g., $\sim$0.13 solar for Complex~C;
\citealt{collins07}), which also argues against their being composed primarily of fountain material.
If the majority of HVCs originate in extragalactic or circumgalactic material, they would travel
great distances through the upper halo, in a relatively low pressure, low density environment
similar to that in our models. As noted above, the pressure gradients may not be large in the upper
halo. Furthermore, clouds that have transverse components to their motion through the halo would
experience less change in ambient pressure in a vertically stratified halo than if they were
traveling vertically downward. Modeling the change in ambient pressure as an HVC falls into the
Galaxy is beyond our current models, but could be incorporated into future simulations.

Our model also assumes a halo temperature of $1 \times 10^6~\K$. X-ray observations indicate the
presence of $\sim$million-degree gas in the halo, although its temperature structure and filling
factor remain uncertain. We did not investigate different ambient temperatures in our suite of
models.  However, in their simulations of 2D plane-parallel mixing layers, \citet{kwak10} found that
increasing the temperature on the hot side of the interface from $1 \times 10^6$ to $3 \times
10^6~\K$ did not have a large effect on the ion column densities for sightlines perpendicular to the
interface.

\citet{kwak10} did not carry out a corresponding simulation with a temperature lower than $1 \times
10^6~\K$.  However, if the halo temperature were much lower than $1 \times 10^6~\K$, the halo's
\rosat\ R2/R1 ratio would be lower than the observed value (e.g., $\mathrm{R}2/\mathrm{R}1 = 0.48$
for a $5 \times 10^5~\K$ plasma, while observational analyses yield $\mathrm{R}2/\mathrm{R}1 \ga
0.5$ for the halo's intrinsic emission; \citealt{snowden00,kuntz00}). If the ambient temperature
were even lower, then the ambient medium would not persist in the upper halo, as it would have a
relatively short cooling time.  For example, for an ambient medium with $T = 3 \times 10^5~\K$ and
$\nH = 1.0 \times 10^{-4}~\pcc$, the cooling time would be $\sim$100~\Myr\ (calculated using the
1993 version of the \citealt{raymond77} code). This time is similar to the lengths of our
simulations.

\subsubsection{\MdotHHVC, \RMW, and Elemental Abundances}
\label{subsubsec:ModelFactors}

When we used Equation~(\ref{eq:Nion}) to calculate the column density of a given ion from our
hydrodynamical simulations, we had to assume values for three important factors: \MdotHHVC, \RMW,
and the elemental abundance (the abundance affects the values of $\Mion(t)$ derived from our
simulations). As noted in Section~\ref{subsec:ColDenPredictions}, the value of \MdotHHVC\ is
uncertain. However, our assumed value of 0.5~\Msolpy\ should be reasonably conservative (in the
sense of tending to lead to underestimates of the predicted column densities), given the large
infall rates associated with individual complexes (e.g., Complex~C, the Smith Cloud, and the
Magellanic Stream). The radius \RMW\ over which the halo ions are distributed is more uncertain. By
choosing $\RMW = 25~\kpc$, we have assumed the ions that result from HVCs interacting with the hot
halo are spread uniformly above the entire stellar disk. However, if these ions tend to be
concentrated toward the center of the Galaxy, which is a reasonable assumption, then our choice of
\RMW\ should also be reasonably conservative.

For the elemental abundances, we used the interstellar values from \citet{wilms00}, which are in
good agreement with recent measurements of solar photospheric abundances
\citep{lodders03,asplund09}.\footnote{Note that the ion masses and column densities in Paper~I were
  calculated using \citet{allen73} abundances, which are 0.14, 0.08, and 0.13~dex larger than the
  \citet{wilms00} abundances for carbon, nitrogen, and oxygen, respectively.} The \citet{wilms00}
interstellar abundances of carbon and oxygen include atoms that are in dust: the gas-phase
abundances of carbon and oxygen in the local ISM are 0.23 and 0.19~dex lower than the
\citeauthor{wilms00} values, respectively \citep{cardelli96,meyer98}. Nitrogen, however, is not
depleted onto dust \citep{meyer97}. Carbon, nitrogen, and oxygen abundances in the halo are more
uncertain, but in the lower halo at least ($|z| \la 2~\kpc$), total abundances appear to be
approximately solar (based on the gas-phase sulfur abundance), and there is less depletion onto dust
than in the disk for elements heavier than oxygen \citep{savage96}. If the carbon, nitrogen, and
oxygen abundances are also approximately solar in the halo, and less depleted onto dust than in the
disk, then our $\Mion(t)$ predictions for \CIV\ and \OVI\ would need to be revised downward by no
more than $\sim$40\%\ to compensate for the possibility that these elements are depleted onto dust,
while our \NV\ predictions would not need revision.

However, it should be noted that varying the abundances would affect the cooling curve used in our
simulations. While the FLASH manual is not explicit, the FLASH cooling curve is likely based on a
cooling curve calculated with \citet{allen73} abundances, as these are the defaults for the
\citet{raymond77} code. Modifying the cooling curve would affect the quantities of high ions
predicted by our simulations, by affecting the amount of gas at optimal temperatures for such ions.
Quantifying the extent of this effect is beyond the scope of this paper. However, while lowering the
abundances would reduce $\Mion(t)$, it would also reduce the cooling rate, meaning that gas would
remain at high-ion-rich temperatures for longer. As a result, lowering the abundances would not
necessarily lead to a commensurate reduction in $\int \Mion(t) dt$, and hence in the predicted column
densities.

\subsection{A Composite Model of the High Ions in the Galactic Halo}
\label{subsec:CompositeModel}

We pointed out in Section~\ref{subsec:ComparisonWithObservations} that the Domain-only column
density predictions in the lower half of Table~\ref{tab:ColDen} are lower limits on the true model
predictions. In the previous subsection, we argued that our choices of \MdotHHVC\ and \RMW\ should
be reasonably conservative, and that adjusting the elemental abundances is unlikely to result in
large adjustments in the predicted column densities. Hence, our statement in
Section~\ref{subsec:ComparisonWithObservations} that our HVC models can account for a significant
fraction of the low-velocity \OVI\ observed in the halo is quite robust.

While the Domain-only predictions from the lower half of Table~\ref{tab:ColDen} are lower limits,
the corresponding Domain + Escaped predictions are generally upper limits on the expected ion column
densities. This is because, for most models, the numbers of low-velocity ions in the material that
has flowed off the domain decreases with time (see Section~\ref{sec:Ions}). Therefore, taking these
lower and upper limits for Model~B (our reference model), we find that HVCs could account for
30--44\%\ of the low-velocity \OVI\ in the halo, but only 12--14\%\ of the low-velocity
\CIV\ (Section~\ref{subsec:ComparisonWithObservations}).

The preceding statement raises the question, why can our model account for a significant fraction
of the \OVI\ in the halo, but not of the \CIV? The answer is likely because our model does not
include photoionization, which can increase the amount of \CIV\ relative to \OVI, due to \CIV's
lower ionization potential. For example, the cooling Galactic fountain model of \citet{shapiro93},
which includes photoionization, predicts $\sim$2--7 times as much \CIV\ relative to \OVI\ as the
cooling fountain model of \citet{edgar86}, which does not. Models that include photoionization
\citep[e.g.,][]{ito88} can explain the enhancement in \CIV\ (and \SiIV, which is not studied in this
paper) relative to \NV\ observed at large $|z|$ \citep{savage97}. However, it should be noted that
the \citet{ito88} model also predicts that \CIV\ should be enhanced relative to \OVI\ at large
$|z|$, which appears not to be the case observationally \citep{savage03}.

As our HVC model does not include photoionization, and therefore tends to significantly underpredict
the amount of \CIV\ in the halo, we are not putting it forward as the only source of high ions in
the halo and as an alternative to other models. Instead, we are pointing out that a complete model
of the Galactic halo should include the contribution of HVCs to the low-velocity high ions,
particularly \OVI. While developing a complete self-consistent model of the high ions in the halo is
beyond the scope of this paper, we will give an example of how different existing models of the halo
high ions could be pieced together. Note that we consider only the total column densities of the
ions, not their $z$ distributions, as we cannot derive the $z$ distributions of the ions from our
current HVC model. However, if the HVCs spend most of their lifetimes in the upper halo (see
Section~\ref{subsubsec:PressureTemperatureHalo}), then the interactions of the HVCs with the ambient
gas may result in a large number of high ions at large distances from the plane.

Our example composite model includes contributions from HVCs (this paper, specifically our reference
model, Model~B), extraplanar SNRs \citep{shelton06}, radiatively cooling Galactic fountains
\citep{shapiro93}, and photoionization by an external radiation field \citep{ito88}. The
contributions from these various components to the observed column densities are presented in
Table~\ref{tab:CompositeModel} (see the footnotes to that table for more details). Note that the
overall normalization of the Galactic fountain component (component~3 in
Table~\ref{tab:CompositeModel}) is essentially a free parameter, as the mass flow rate in the
fountain is not well known. We therefore rescaled the predictions from \citet{shapiro93} so that
this component would account for all of the \OVI\ not accounted for by HVCs and SNRs (this rescaling
amounted to multiplying their predictions by 0.24--0.38, depending on the particular model in their
Table~1). Hence, our model accounts for all of the observed \OVI\ by design.

\begin{deluxetable}{lccc}
\tablewidth{0pt}
\tablecaption{Composite Model of Low-Velocity High Ions in Halo\label{tab:CompositeModel}}
\tablehead{
                                                              & \colhead{\CIV}        & \colhead{\NV}         & \colhead{\OVI}
}

\startdata
\phantom{(1)} Observed\tablenotemark{a} ($10^{13}~\pcmsq$)     & 10                    & 2.5                   & 15.5          \\
(1)           HVCs\tablenotemark{b}                           & 13\%                  & 22\%                  & 37\%          \\
(2)           Extraplanar SNRs\tablenotemark{c}               & 6\%                   & 12\%                  & 17\%          \\
(3)           Galactic fountains\tablenotemark{d}             & 23--37\%              & 14--21\%              & 46\%          \\
(4)           Photoionization by \\
\phantom{(4)} external radiation field\tablenotemark{e}       & 39\%                  & \nodata               & \nodata       \\
\phantom{(1)} Total                                           & 81--95\%              & 47--55\%              & 100\%\tablenotemark{f}
\enddata

\tablecomments{Each model component's contribution is expressed as a percentage of the observed column density, in the first row.
  Any apparent discrepancies between the individual components' contributions and the totals are due to rounding.}
\tablenotetext{a}{\CIV: \citet{savage09}; \NV: \citet{savage97}; \OVI: \citet{savage03}, after removing contribution from the Local Bubble \citep{oegerle05}.}
\tablenotetext{b}{Predictions from Model~B (our reference model); specifically, the average of the Domain-only and Domain + Escaped predictions
  from the lower half of Table~\ref{tab:ColDen}.}
\tablenotetext{c}{\citet{shelton06}, Table~8. We have chosen the model with the median \OVI\ prediction (case~1, drag coefficient~1).}
\tablenotetext{d}{\citet{shapiro93}, Table~1, rescaled such that this component accounts for all of the \OVI\ not already accounted for by HVCs and SNRs. This model includes the effects of self-photoionization.}
\tablenotetext{e}{\citet{ito88}, Figure~3(a). We assume that all of the \CIV\ in this figure is due to the photoionized component of their model, and that this component produces negligible \NV\ and \OVI.}
\tablenotetext{f}{Our composite model reproduces 100\%\ of the \OVI\ by design.}
\end{deluxetable}

In our composite model, HVCs, SNRs, and Galactic fountains (components~1--3) can account for
approximately half of the observed \CIV\ and \NV. The Galactic fountain model of \citet{shapiro93}
includes the effects of photoionization by the cooling gas's own radiation, but not by an external
radiation field. Photoionization by the extragalactic background, or by radiation from hot stars
escaping from the disk, may be able to account for the shortfall in \CIV.  For example, the hybrid
model of \citet{ito88} includes a photoionized component with $T = 10^{4.0}~\K$, which could account
for $\approx$40\%\ of the observed \CIV\ (component~4 in Table~\ref{tab:CompositeModel}); i.e., most
or nearly all of the observed \CIV\ is now accounted for. However, this photoionized component is
unlikely to be able to account for the additional \NV, due to \NV's higher ionization potential.

The shortfall in \NV\ in our composite model is due to the fact that the various components of our
composite model all underpredict the \NV\ column density relative to the \OVI\ column density.  The
predicted ion ratios for HVCs (this paper), extraplanar SNRs \citep{shelton06}, and Galactic
fountains \citep{edgar86,shapiro93} are all in the range $\log \left[ \NNV / \NOVI \right] \approx
-1.3$ to $-1.0$. In comparison, the observed value is $\log \left[ \NNV / \NOVI \right] = -0.8$ (see
Table~\ref{tab:CompositeModel}). Some additional source of gas with $T \approx 2 \times 10^5~\K$
would be needed to increase the \NV\ column density relative to the \CIV\ and \OVI\ values, although
what this source could be is not obvious. Plane-parallel turbulent mixing layer models tend to give
higher \NV-to-\OVI\ ratios ($\log \left[ \NNV / \NOVI \right] = -0.9$ to $-0.4$;
\citealt{slavin93a,kwak10}), but as the composite model already includes mixing between HVC and
ambient gas, it is not clear how more turbulent mixing layers could be added to the model.

\section{SUMMARY}
\label{sec:Summary}

We have presented further results from our NEI hydrodynamical simulations of HVCs (Paper~I). In this
paper, we concentrated on the low-velocity high ions that result when an HVC passes through the hot
halo. These high ions arise from the mixing of cool cloud gas with hot ambient gas. Initially, this
mixed high-ion-rich gas is located near the HVC, and travels with HVC-like velocities, but later it
falls behind the HVC and slows to ISM-like velocities, while retaining its high ion content.

We examined a suite of seven models, covering different cloud velocities, cloud densities, cloud
density profiles, and cloud sizes. In general, the cloud velocity and density profile have little
effect on the masses of the high ions that result. Larger or denser clouds result in greater masses
of high ions, as one would qualitatively expect. An important result is that, except for the fastest
model HVC ($|v| = 300~\kmps$), the quantities of low-velocity ions are generally larger than the
quantities of high-velocity ions. This result suggests that HVCs could be an important source of
low-velocity high ions in the halo.

We examined this suggestion more quantitatively, using the HVC infall rate to estimate the average
column densities of low-velocity high ions in the halo due to HVCs. After accounting for the
contribution from ions at times beyond the ends of our simulations, and being conservative regarding
the HVC infall rate, we find that our models can account for $\ga$30\%\ of the \OVI\ column density
observed in the halo. This implies that the collisionally ionized gas in material shed by HVCs is a
significant source of the low-velocity \OVI\ observed in the halo. In contrast, such gas is probably
not a significant source of low-velocity \CIV: our reference model can account for only
12--14\%\ of the observed column density. This is probably because the observed \CIV\ is likely
affected by photoionization, which our model does not include.

We used the predictions of our HVC model in a simple composite model of the low-velocity high ions
in the halo, in which we combined the contributions from HVCs, extraplanar SNRs, radiatively cooling
fountain gas, and photoionization from an external radiation field. By design, this model accounted
for all of the observed \OVI. We found that the model could account for most or all of the observed
\CIV, but only about half of the observed \NV. It is not obvious what the source of the additional
\NV\ could be. Although this composite model was constructed in a relatively simple way, and failed
to account for all the observed ions, we emphasize the point that any complete model of the high
ions in the halo should include the contributions from HVCs, particularly to the column density of
low-velocity \OVI.

\acknowledgements

The software used in this work was in part developed by the DOE-supported ASC/Alliance Center for
Astrophysical Thermonuclear Flashes at the University of Chicago. The simulations were performed at
the Research Computing Center (RCC) of the University of Georgia. We acknowledge use of the R
software package \citep{R}. This work was supported by NASA grant NNX09AD13G, awarded through the
Astrophysics Theory and Fundamental Physics Program. DBH acknowledges funding from NASA grant
NNX08AJ47G, awarded through the Astrophysics Data Analysis Program.

\begin{appendix}

\section{A Simple Analytical Model of a Spherical HVC}

Here we describe a simple analytical model of a spherical HVC that loses mass at a rate proportional
to its surface area (introduced in Section~3.3.3 of Paper~I). This model is helpful for comparing
the results from hydrodynamical simulations of different-sized HVCs (see
Sections~\ref{subsec:DifferentParameters} and \ref{subsec:ModelComparison}).

We assume that the mass of low-velocity high ions of a given type, $\Mion$, increases with time at
a rate that is proportional to the rate at which the HVC sheds its mass, and thus to the HVC's surface
area, i.e., $d\Mion/dt = \alpha r^2(t)$. In this simple model, the cloud radius decreases linearly
with time from its initial value, $r_0$ (see Paper~I), and so we can rewrite the mass loss rate as
$d\Mion/dt = \alpha (r_0 - kt)^2$.  Note that the constants $\alpha$ and $k$ are unimportant, as
long as they are independent of $r_0$. Note also that we are ignoring the fact that the ions of
interest will subsequently recombine or ionize (i.e., there should be an additional, negative term
on the right-hand side of the expression for $d\Mion/dt$).

We integrate $d\Mion/dt$ with respect to $t$, with the boundary condition $\Mion = 0$ at $t=0$. In order to
compare different-sized clouds at equivalent stages of their evolution, we
express the solution as a function of the rescaled time, $t/r_0$, obtaining
\begin{equation}
  \Mion(t/r_0) = \frac{\alpha r_0^3}{3k} \left[ 1 - \left\{ 1 - k \left( \frac{t}{r_0} \right) \right\}^3 \right].
  \label{eq:Mion}
\end{equation}
Hence, we expect that
\begin{equation}
  \Mion(t/r_0) \propto r_0^3 \propto \MHVCinit.
  \label{eq:Mion2}
\end{equation}
We use this result in Section~\ref{subsec:DifferentParameters}. Note that this result is only valid
if $t < r_0/k$ and if the rate at which the ions of interest subsequently recombine or ionize is
negligible.

We can also use this model to compare the ion column densities expected from different-sized
clouds. If the end time, \tf, for the integral in Equation~(\ref{eq:Nion}) is chosen such that $\tf
/ r_0$ is the same for different-sized clouds, then we expect that
\begin{eqnarray}
  \Nbarion &\propto& \frac{\int_{0}^{\tf} \Mion dt}{\MHVCHIinit}                   \nonumber \\
           &   =   & r_0 \frac{\int_{0}^{\tf/r_0} \Mion(t/r_0) d(t/r_0)}{\MHVCHIinit},  \nonumber \\
           &\propto& r_0,
  \label{eq:Nion_r0}
\end{eqnarray}
where, for the final step, we have used the fact that $\Mion(t/r_0) \propto \MHVCHIinit$ in this
simple model (Equation~(\ref{eq:Mion2})).  We use this result in
Section~\ref{subsec:ModelComparison}. Again, note that this result is only valid if the rate at
which the ions of interest ionize or recombine is negligible.

\end{appendix}

\bibliographystyle{apj}
\bibliography{references}

\end{document}